\documentclass[12pt, draftcls, peerreview]{IEEEtran}
\usepackage{mathrsfs}
\usepackage{graphicx}
\usepackage{float}
\usepackage{booktabs}
\usepackage{hyperref}
\usepackage{cite}

\usepackage{enumitem}
\usepackage{color}
\usepackage{ulem}
\usepackage{psfrag}
\usepackage{subfigure}
\usepackage{amssymb}
\usepackage{amsthm}
\usepackage{setspace}
\usepackage{epsfig}
\usepackage{pifont}
\usepackage{amsmath}
\usepackage{array}
\newcommand{\PreserveBackslash}[1]{\let\temp=\\#1\let\\=\temp}
\newcolumntype{C}[1]{>{\PreserveBackslash\centering}p{#1}}
\newcolumntype{R}[1]{>{\PreserveBackslash\raggedleft}p{#1}}
\newcolumntype{L}[1]{>{\PreserveBackslash\raggedright}p{#1}}
\usepackage{multicol}
\usepackage{multirow}
\usepackage{diagbox}
\usepackage{pifont}
\usepackage{indentfirst}
\usepackage{amsfonts}
\usepackage{algorithm}
\usepackage{algorithmicx}
\usepackage{algpseudocode}
\usepackage{fancyhdr}
\usepackage{amscd}
\usepackage[cmyk]{xcolor}
\usepackage{bm}
\usepackage{fancyhdr}
\usepackage{enumerate}
\setlist{itemsep=0pt,parsep=0pt}

\renewcommand{\arraystretch}{1.2}
\newtheorem{proposition}{Proposition}

\newtheorem{remark}{Remark}
\newcommand{\tabincell}[2]{\begin{tabular}{@{}#1@{}}#2\end{tabular}}
\renewcommand{\arraystretch}{1.1} 
\allowdisplaybreaks[4]
\IEEEoverridecommandlockouts

\renewcommand{\arraystretch}{0.9}	
\usepackage{caption}
\begin{document}
	\title{A Model-based GNN for Learning Precoding}
		
\author{
	\IEEEauthorblockN{Jia Guo and Chenyang Yang}
	\thanks{If this paper is accepted, we will publicize our codes on GitHub.}
%
\vspace{-5mm}}
	\maketitle
	\setcounter{page}{1}
	\thispagestyle{headings}
	
\begin{abstract}\vspace{-3mm}
Learning precoding policies with neural networks enables low complexity online implementation, robustness to channel impairments, and joint optimization with channel acquisition. However, existing neural networks suffer from high training complexity and poor generalization ability when they are used to learn to optimize precoding for mitigating multi-user interference. This impedes their use in practical systems where the number of users is time-varying. In this paper, we propose a graph neural network (GNN) to learn precoding policies by harnessing both the mathematical model and the property of the policies. We first show that a vanilla GNN cannot well-learn pseudo-inverse of channel matrix when the numbers of antennas and users are large, and is not generalizable to unseen numbers of users.
Then, we design a GNN by resorting to the Taylor's expansion of matrix pseudo-inverse, which allows for capturing the importance of the neighbored edges to be aggregated that is crucial for learning precoding policies efficiently.
Simulation results show that the proposed GNN can well learn spectral efficient and energy efficient precoding policies in single- and multi-cell multi-user multi-antenna systems with low training complexity, and can be well generalized to the numbers of users.

\begin{IEEEkeywords}\vspace{-1mm}
	Graph neural network, model, precoding, matrix pseudo-inverse, permutation equivariance
\end{IEEEkeywords}
\end{abstract}

\vspace{-3mm}
\section{Introduction}
Optimizing precoding is critical for boosting the spectral efficiency (SE) \cite{WMMSE2011Shi} and energy efficiency (EE) \cite{LY_TVT_2015} of multi-user multi-input-multi-output (MU-MIMO) systems. Although a variety of numerical algorithms have been proposed to solve the non-convex problems for optimizing precoding, their computational complexities are high when the number of users is large, and their solutions are sensitive to the impairments such as channel estimation errors. This hinders their applications in 5G and 6G systems with real-time requirements.

Recently, deep neural networks (DNNs) have been introduced to optimize wireless policies such as power control \cite{L2O}, link scheduling \cite{lee2020wireless} and precoding\cite{FNN_mUE_DigPre_TVT2021_Kong}. With deep learning, precoding and channel estimation can be optimized jointly \cite{GNN_mUE_RIS_JSAC_2021_Yu}, and the learned policies can be implemented with low running time \cite{L2O}.
Despite these advantages, existing DNNs suffer from high complexity for training a large number of free parameters with massive data sets and long training time \cite{ZBC_WCNC}, and are not ensured generalizable to every impacting factor
such as the channel distribution \cite{L2ContinueOpt} and the number of users.
Since wireless systems are operated in open and  dynamic environments, the DNNs for learning wireless policies have to be re-trained whenever the non-generalizable impacting factors change significantly. This calls for the boosting of learning efficiency, by improving the generalization ability of DNNs to decrease the re-training frequency, and by reducing the complexity for training the DNNs \cite{SCJ20ranking}.

One way of improving learning efficiency is to incorporate the mathematical property of the policies when designing DNN structures, which helps reduce the hypothesis space of the DNNs. For example, graph neural networks (GNNs) can harness the permutation equivariance (PE) property  existed in many wireless policies  \cite{Eisen2020, SCJ20ranking}, which have been used to  learn power allocation \cite{Eisen2020, GJ_TWC_GNN} and precoding \cite{ZBC_WCNC} policies.
While the training complexity can be reduced remarkably compared to the non-structural fully connected neural networks (FNNs),
 the GNNs still require a large training set and cannot be generalized to unseen numbers of users when they are used for learning to optimize precoding in MU-MIMO systems.

Another way is to integrate mathematical models with DNNs, which can be an iterative algorithm or the structure of an optimal solution \cite{shlezinger2020model}. For example, deep unfolding, which is a popular model-driven learning framework, imitates the iterative procedure of an algorithm by the layers of DNNs and only learns several operations or parameters \cite{DeepUnfold_WMMSE_TWC_2021}. Then, the mappings that need to be learned  are simplified, and hence the learning performance can be improved.

\vspace{-4mm}\subsection{Related Works}
\vspace{-1mm}
\subsubsection{Learning to optimize precoding policies}
Deep learning has been introduced to optimize precoding  in MU-MIMO systems recently, which enables to reduce the online inference time \cite{FNN_mUE_DigPre_WCL_2020_Kim},  jointly optimize with channel acquisition \cite{GNN_mUE_RIS_JSAC_2021_Yu}, and improve robustness against channel estimation errors \cite{CNN_sUE_HyPre_TCOM2022_Liu}. FNNs were used to learn precoding policies in \cite{FNN_mUE_DigPre_TVT2021_Kong, FNN_mUE_DigPre_WCL_2020_Kim},  where the FNN was trained in an unsupervised manner to avoid generating labeled samples in \cite{FNN_mUE_DigPre_WCL_2020_Kim}. To extract spatial correlations from channel matrices or to reduce trainable parameters, convolutional neural networks (CNNs) were introduced to optimize hybrid precoding in \cite{CNN_sUE_HyPre_TCOM2022_Liu, CNN_sUE_HyPre_TWC2020_Peken} for millimeter-wave (mmWave) communication systems.
To improve training efficiency, GNNs were designed in \cite{ZBC_WCNC, shen2019graph} to optimize precoding. In \cite{GNN_mUE_RIS_JSAC_2021_Yu}, a GNN was designed to learn beamforming and reflective pattern in intelligent reflecting surface (IRS) systems.  In \cite{FNN_mUE_DigPre_TVT2021_Kong, GNN_mUE_RIS_JSAC_2021_Yu, DL_ChlEst_Precod_TWC_2021},
channel estimation (or pilot design) and precoding were jointly optimized.
All these works only  learn the SE-maximal precoding policies, except that the precoding for maximizing the minimal data rate was also considered  in \cite{GNN_mUE_RIS_JSAC_2021_Yu}. The coordinated beamforming was learned in \cite{DL_mBS_DigPre_GC2020_Lee}.

Without leveraging prior knowledge, the DNNs in most prior works need high training complexity even for small-scale problems with a few antennas and users. While the PE properties of the precoding policies have been harnessed by GNNs in \cite{ZBC_WCNC, shen2019graph, GNN_mUE_RIS_JSAC_2021_Yu}, the learning performance degrades when the signal-to-noise-ratio (SNR) is high and the number of users is large.


\subsubsection{Model-driven deep learning}
Deep unfolding was first proposed in \cite{DU_SparseCode_LeCun_2010}, where a neural network resembles the iteration process of a shrinkage-thresholding algorithm.
This approach has been introduced into wireless communications recently to improve the learning efficiency. A weighted minimum mean-square-error (WMMSE) algorithm \cite{WMMSE2011Shi} was unfolded to optimize a SE-maximal precoding in \cite{DeepUnfold_WMMSE_TWC_2021}, aimed to reduce the computational complexity of matrix inverse operation in the algorithm. The WMMSE algorithm was unfolded in \cite{DeepUnfold_WMMSE_Arindam_2020} to optimize SE-maximal power allocation, aiming to accelerate the convergence of the algorithm.
Signal detectors were optimized by unfolding an orthogonal approximate message passing algorithm in \cite{DeepUnfold_Learn_to_Detect_TSP_2019} and gradient descent-based
algorithms in \cite{DeepUnfold_MIMO_Detect_TSP_2020}, where only a few parameters such as the step size were learned from data. Instead of solving a specific problem, a generalized benders decomposition algorithm was considered in \cite{learnGBD_TWC_2020} to optimize the mixed integer nonlinear programming (MINLP) problem, where neural networks were used to learn the decision of whether a constraint obtained by solving the primal problem should be added into the master problem.
Despite that the training complexity can be reduced, some deep unfolding neural networks (e.g., the one in \cite{DeepUnfold_WMMSE_TWC_2021}) are still hard to be implemented in real-time in large-scale wireless systems. Moreover, these methods are algorithm-specific or problem-specific. For example, the WMMSE algorithm cannot be used to optimize EE-maximal policies, and the
MINLP problem solved by the generalized benders decomposition algorithm should be convex for the continuous variables.

Another model-driven approach for precoding was proposed in \cite{DL_Beamforming_model_TCOM2020, DL_Beamforming_model_TVT_2020}, which exploits the structure of optimal solution such that only the power allocation needs to be learned. In \cite{DL_BF_StrucZF_TWC2022}, by assuming that the optimal solution structure of channel estimation and SE-maximal precoding are respectively composite functions of the least-square estimator and zero-forcing beamforming (ZFBF), only the two composite functions were learned to jointly optimize channel estimation and precoding. {A similar approach was proposed in \cite{DL_Precod_InvExpan_CL2022}, where a neural network was used to learn a function of minimum-mean-square-error beamforming (also referred to as regularized zero-forcing beamforming (R-ZFBF) in the literature) for maximizing SE}. To reduce the complexity of computing matrix inverse, {it is approximated by Taylor's expansion as in \cite{DeepUnfold_WMMSE_TWC_2021} and then input into a DNN for learning precoding.}
While the learning performance can be improved or the training complexity can be decreased, most prior works except \cite{DL_BF_StrucZF_TWC2022} do not leverage the PE property in designing the neural networks, which makes them not generalizable to problem scales. Moreover, this kind of methods highly depends on the optimal solution structure,  which is problem-specific.

In summary, how to design model-based DNNs for learning precoding policies from various optimization problems efficiently is still open.

\vspace{-4mm}\subsection{Motivations and Contributions}\vspace{-2mm}
Based on the fact that the pseudo-inverse of a matrix is explicitly or implicitly required in precoding for avoiding multi-user interference in MU-MIMO systems and is hard to learn due to the non-continuity of the inverse function, in this paper we propose a GNN to learn pseudo-inverse-related precoding policies by harnessing the first-order Taylor's expansion formula. The GNN can satisfy the PE property of the policies. We strive to improve the learning efficiency of the GNN by reducing training complexity and allowing generalizability. In the literature of learning to optimize precoding, the generalization ability of the DNNs to channel distribution has been evaluated in \cite{DL_ChlEst_Precod_TWC_2021}, but the generalization ability to larger problem scales is only considered in \cite{GNN_mUE_RIS_JSAC_2021_Yu}. Since the traffic load in a cell changes quickly, we concentrate on the generalizability of the GNN to the problem scale, i.e., a  well-trained GNN can be used for inference in scenarios with unseen numbers of users without the need for re-training.
To facilitate analyses and for easy exposition, we take a simple but representative precoding in multiple-input-single-output (MISO) system as an example.
The major contributions are summarized as follows.
\begin{itemize}
	\item We use first-order Taylor's expansion to approximate matrix pseudo-inverse iteratively, which can be regarded as the update equation of a perfectly-trained GNN to learn a ZFBF policy. We then introduce trainable parameters into the update equation to design a Model-GNN, which play the role of adjusting the power and direction of the precoding vectors. The update equation of the Model-GNN enables aggregating information of edges with different weights, which is critical for improving the generalization ability of GNN to problem scales.  
	Different from the deep unfolding method \cite{DeepUnfold_WMMSE_TWC_2021}, we do not mimic a specific iterative algorithm for optimizing precoding. Different from \cite{DL_Beamforming_model_TCOM2020, DL_Beamforming_model_TVT_2020, DL_BF_StrucZF_TWC2022}, we do not assume any structure of the optimal solution for precoding.
		
	\item We prove that when the numbers of users and antennas are large, a vanilla GNN with commonly used pooling, combination and processing functions cannot learn the precoding policies unless the policy is an element-wise function. By comparing the update equations of the proposed GNN and the vanilla GNN, we explain why the vanilla GNN for learning ZFBF policy cannot be generalized to the number of users.

	\item We extend the Model-GNN for learning to optimize coordinated beamforming in multi-cell interference systems, to demonstrate that the GNN is applicable for learning various policies with different objectives and constraints in different scenarios. Simulation results show that the proposed Model-GNN can achieve good performance for learning SE-maximal and EE-maximal precoding policies in different settings, is generalizable to the number of users, and requires lower training complexity to achieve an expected performance.
\end{itemize}

\emph{Notations}: ${\sf Tr}(\cdot)$ denotes trace of a matrix, $\|\cdot\|$ denotes two-norm, ${\sf Re}(\cdot)$ and ${\sf Im}(\cdot)$ are respectively the real and imaginary part of a complex value, $f'(\cdot)$ is the derivative of $f(\cdot)$, ${\bf X}=[x_{ij}]^{K\times N}$ denotes a $K\times N$ matrix, and its element in the $i$th row and $j$th column is $x_{ij}$, ${\bf I}_K$ denotes an $K\times K$ identity matrix.

The rest of the paper is organized as follows. In section \ref{sec: system model}, we introduce precoding policies and a vanilla GNN to learn the policies. Then, we prove that the GNN cannot well-learn matrix inverse-based precoding policies when the problem scale is large. In section \ref{sec: model-gnn}, we design a GNN  to incorporate the Taylor's expansion formula of matrix pseudo-inverse. In section \ref{sec: simulations}, we provide simulation results to compare the performance of the GNNs with existing model-based DNNs, and evaluate the generalization ability of the GNNs to the number of users. In section \ref{sec: conclusions}, we provide the conclusion remarks.

\vspace{-4mm}\section{Learning to Optimize Precoding Policy with a GNN}\label{sec: system model}\vspace{-2mm}
In this section, we consider the precoding problems with closed-form solutions to show the role of matrix inverse in computing precoding. We provide the PE property of the resulting precoding policies, and show that a vanilla GNN whose input-output relationship satisfies the PE property cannot  learn the matrix-inverse-based policies when the problem scale is very large.
\subsection{Precoding Policy: Role of Matrix Inverse}\vspace{-1mm}
Consider a downlink MISO system, where a base station (BS) equipped with $N$ antennas transmits to $K$ users each with a single antenna.
The precoding matrix can be optimized to maximize an objective function subject to a set of constraints.

For easy exposition, we first consider a widely-studied problem, which optimizes precoding to maximize SE under the power constraint at the BS,\vspace{-1mm}
\begin{subequations}\label{P1: SE max}
	\begin{align}
	{\emph {P1}}:~~~\max_{\bf V}~~ & \sum_{k=1}^K R_k({\bf H, V}) \triangleq \log_2\Bigg(1 + \frac{|{\bf h}_k^{\sf H}{\bf v}_k|^2}{\sum_{j=1,j\neq k}^K|{\bf h}_k^{\sf H}{\bf v}_j|^2 + \sigma_0^2}\Bigg) \label{P1: obj} \\
	{\rm s.t.}~~ & {\sf Tr}({\bf V}^{\sf H}{\bf V}) \leq P_{\max}, \label{P1: constraint} \vspace{-1mm}
	\end{align}
\end{subequations}
where ${\bf H}=[{\bf h}_1,\cdots,{\bf h}_K]\in{\mathbb C}^{N\times K}$, ${\bf h}_k=[h_{1k}, \cdots, h_{Nk}]^{\sf T}$  is the channel vector of the $k$th user (denoted as UE$_k$), ${\bf V}=[{\bf v}_1,\cdots,{\bf v}_K]\in{\mathbb C}^{N\times K}$,  ${\bf v}_k=[v_{1k}, \cdots, v_{Nk}]^{\sf T}$ is the precoding vector for UE$_k$, 
which is composed of a beamforming direction ${\bf v}_k/\|{\bf v}_k\|$ and a power allocated to UE$_k$, i.e., $\|{\bf v}_k\|^2$, 
$P_{\max}$ is the maximal transmit power of the BS, and $\sigma_0^2$ is the noise power.

The optimal solution of problem \emph{P1} has the following structure \cite{Precod_Opt_Structure},\vspace{-2mm}
\begin{equation}\label{eq: opt precod struc}
{\bf V}^{\star}
 =\mathbf{H}\left(\boldsymbol{\Lambda}\mathbf{H}^{\sf H}\mathbf{H}+\sigma_0^2\mathbf{I}_K\right)^{-1}{\mathbf{T}}^\frac{1}{2}
 \triangleq f_2\Big(\big(f_1({\bf H})\big)^{-1}, {\bf H}\Big),\vspace{-2mm}
\end{equation}
where
$f_1({\bf H}) \triangleq \boldsymbol{\Lambda}\mathbf{H}^\mathsf{H}\mathbf{H} + \sigma_0^2\mathbf{I}_K$, $\bm \Lambda$
is a diagonal matrix of Lagrange multipliers whose trace is $P_{\max}$, and ${\bf T}$ is a diagonal matrix of the allocated powers to users  whose trace is also $P_{\max}$.

As shown in  \cite{Precod_Opt_Structure},  if both the objective function and power constraint of a precoding problem  in MISO systems strictly increase with the signal-to-interference-and-noise ratio (SINR) of each user,  then the optimal solution of the problem will have the structure in \eqref{eq: opt precod struc}.

When SNR is very high, the optimal precoding approaches to ZFBF (i.e., ${\bf V}_{\sf ZF}\triangleq {\bf H}({\bf H}^{\sf H}{\bf H})^{-1}$) followed by power allocation. When SNR is very low, the optimal precoding approaches to maximum-ratio-transmission (MRT)  (i.e., ${\bf V}_{\sf MRT}\triangleq {\bf H}$)  followed by power allocation \cite{Precod_Opt_Structure}. When SNR is moderate, the optimal precoding has similar structure with another commonly used precoding,  R-ZFBF (i.e., ${\bf V}_{\sf RZF}\triangleq {\bf H}({\bf H}^{\sf H}{\bf H}+\frac{K\sigma_0^2}{P_{\max}}{\bf I}_K)^{-1}$)
followed by power allocation.


A precoding policy is the mapping from environmental parameters to a precoding matrix. For notational simplicity, we only consider the channel as the environmental parameter. Hence, the policy is denoted as ${\bf V}^{\star}=f({\bf H})$. Different problems yield different policies.
From  \eqref{eq: opt precod struc}, we can see that several optimal precoding policies consist of matrix (pseudo-) inversion operation, which is essential to avoid multi-user interference for all precoding policies \cite{serbetli2004transceiver}.

\vspace{-4mm}\subsection{Property of the Precoding Policies and a Vanilla GNN to Learn the Policies} \label{sec: edge gnn} \vspace{-1mm}
When the order of users and the order of antennas are respectively changed by the permutation matrices ${\bf \Pi}_{\sf UE}$  and ${\bf \Pi}_{\sf AN}$, the channel and precoding matrices change accordingly but the objective function and the constraints of problem \emph{P1} remain the same.
 This indicates that the precoding policies satisfy the following PE property,
${\bf \Pi}_{\sf AN}^{\sf T}{\bf V}^{\star}{\bf \Pi}_{\sf UE}=f({\bf \Pi}_{\sf AN}^{\sf T}{\bf H}{\bf \Pi}_{\sf UE})$. It is easy to show that the ZFBF, MRT and R-ZFBF policies also satisfy the  PE property.

\begin{figure}[!htb]
	\centering
	\begin{minipage}[t]{0.35\linewidth}	
		\subfigure[Vanilla-GNN]{
			\includegraphics[width=\textwidth]{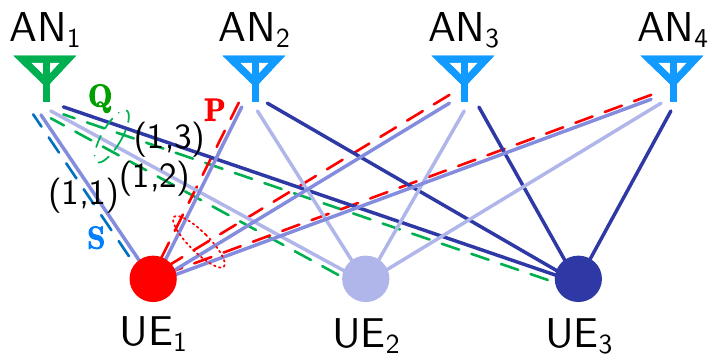}}
	\end{minipage} \hspace{10mm}
	\begin{minipage}[t]{0.35\linewidth}	
		\subfigure[Model-GNN]{
			\includegraphics[width=\textwidth]{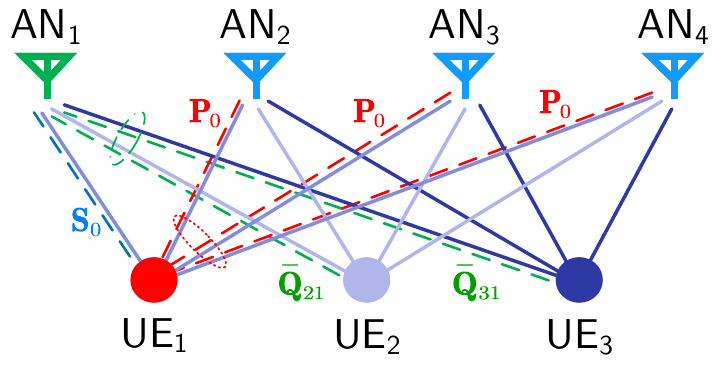}}
	\end{minipage}
	\vspace{-4mm}
	\caption{Example of precoding graph when $N=4$, $K=3$ and updating ${\bf d}_{11}^{(\ell)}$ with (a) Vanilla-GNN and (b) Model-GNN (to be introduced later), where the solid lines are wireless links and the dashed lines indicate the trainable weights. ${\bf d}_{11}^{(\ell)}$ is obtained by first aggregating information from the edges that are neighbored to edge $(1,1)$ by ${\sf AN}_1$ (i.e., edge $(1,2)$ and edge $(1,3)$) and by ${\sf UE}_1$ (i.e., edges in the red circle) and then combining with the information of edge $(1, 1)$.  }\label{fig-precoding-graph}\vspace{-1mm}
\end{figure}

These precoding policies can be learned over a graph, which is composed of two types of vertexes (i.e., antennas at the BS and users) and edges (i.e., the links between the vertexes). The vertexes have no feature. Denote the edge between the $n$th antenna (denoted as AN$_n$) and UE$_k$ as edge $(n, k)$. The feature of each edge (say edge $(n,k)$) is $h_{nk}$, and the action of edge $(n,k)$ is $v_{nk}$. Hence, learning a precoding policy with GNN is to learn the actions from the features of the graph, which is referred to as \emph{precoding graph}. When learning the policies from different problems, only the loss functions for training the GNN differ. As  illustrated in Fig. \ref{fig-precoding-graph}, edge $(1, 1)$ is connected to the same vertex (i.e., AN$_1$) with edge $(1, 2)$ and edge $(1, 3)$. We refer to edge $(1, 2)$ and edge $(1, 3)$ as the \emph{neighbored edges} of edge $(1, 1)$ by AN$_1$.

Since both the features and actions are defined on edges, an edge-GNN is used to learn the precoding policy  as in \cite{ZBC_WCNC}, where each layer of the GNN outputs a representation of each edge. The output of edge $(n,k)$ in the $\ell$-th layer, ${\bf d}_{nk}^{(\ell)}$, is obtained from the following \emph{update equation},\vspace{-1mm}
\vspace{-4mm}\begin{equation}\label{eq: edge gnn upd}
{\bf d}_{nk}^{(\ell)} = {\sf CB}\Bigg({\bf d}_{nk}^{(\ell-1)}, \underbrace{{\sf PL}_{\sf u}~_{i=1,i\neq n}^N\Big(q_{\sf u}({\bf d}_{ik}^{(\ell-1)}, {\bf P}^{(\ell)})\Big)}_{(a)}, \underbrace{{\sf PL}_{\sf b}~_{j=1,j\neq k}^K\Big(q_{\sf b}({\bf d}_{nj}^{(\ell-1)}, {\bf Q}^{(\ell)})\Big)}_{(b)}, {\bf S}^{(\ell)}\Bigg),\vspace{-1mm}
\end{equation}
where terms $(a)$ and $(b)$ are the aggregated outputs of the neighbored edges of edge $(n,k)$ by UE$_k$ and AN$_n$, respectively, ${\sf PL}_{\sf u}(\cdot)$ and ${\sf PL}_{\sf b}(\cdot)$ denote the \emph{pooling functions} that are used for aggregating the outputs of the neighbored edges of edge $(n,k)$ by UE$_k$ and AN$_n$, respectively, $q_{\sf u}(\cdot, {\bf P}^{(\ell)})$ and $q_{\sf b}(\cdot, {\bf Q}^{(\ell)})$ denote the \emph{processing functions} with trainable parameters ${\bf P}^{(\ell)}$ and ${\bf Q}^{(\ell)}$ that are used for extracting useful information from the neighbored edges of edge $(n,k)$ by UE$_k$ and AN$_n$, respectively, and ${\sf CB}(\cdot)$ denotes a \emph{combination function} with trainable parameters ${\bf S}^{(\ell)}$ that combines the aggregated outputs with the output of edge $(n,k)$ in the $(\ell-1)$th layer.

To ensure that the permutation of the vertexes with the same type does not affect the outputs of GNN, each processing function should be identical for all the edges, and the pooling functions should satisfy commutative law (e.g., summation and maximization).

Without loss of generality, $q_{\sf u}(\cdot)$ and $q_{\sf b}(\cdot)$ in \eqref{eq: edge gnn upd} can be selected from the same family of functions (e.g., linear functions) with different trainable parameters ${\bf P}^{(\ell)}, {\bf Q}^{(\ell)}$, and ${\sf PL}_{\sf u}(\cdot)$ and ${\sf PL}_{\sf b}(\cdot)$ can also be the same function (e.g., summation function). Then, the subscripts ${\sf u}$ and ${\sf b}$ can be omitted.
For notational simplicity, we assume that the trainable parameters are identical in all the layers, then the superscript $(\ell)$ of the three matrices ${\bf P}^{(\ell)}, {\bf Q}^{(\ell)},{\bf S}^{(\ell)}$ in \eqref{eq: edge gnn upd} can be omitted.

When $q(\cdot)$ is a linear function and ${\sf CB}(\cdot)$ is a linear function cascaded with an activation function $\sigma(\cdot)$, which are commonly used in the literature \cite{GJ_TWC_GNN,ZBC_WCNC,lee2020wireless}, \eqref{eq: edge gnn upd} is degenerated to,\vspace{-1mm}
\begin{equation}\label{eq: edge gnn upd1}
{\bf d}_{nk}^{(\ell)} = \sigma\Big({\bf S}{\bf d}_{nk}^{(\ell-1)} + {\sf PL}_{i=1,i\neq n}^N\big({\bf P}{\bf d}_{ik}^{(\ell-1)}\big) + {\sf PL}_{j=1,j\neq k}^K\big({\bf Q}{\bf d}_{nj}^{(\ell-1)}\big) \Big).\vspace{-1mm}
\end{equation}

When the edge-GNN with update equation in \eqref{eq: edge gnn upd1} is used to learn the precoding policy, its input is ${\bf d}_{nk}^{(0)}=[{\sf Re}(h_{nk}), {\sf Im}(h_{nk})]$, and its output is ${\bf d}_{nk}^{(L)}=[{\sf Re}(\hat{v}_{nk}), {\sf Im}(\hat{v}_{nk})]$, where $\hat{v}_{nk}$ is the element in the $n$th row and $k$th column of the learned precoding matrix $\hat{\bf V}=[\hat{v}_{nk}]^{N\times K}$ and $(L+1)$ is the number of layers, i.e., there is an input layer (the $0$th layer), an output layer (the $L$th layer) and $(L-1)$ hidden layers.
The input-output relationship is denoted as $\hat{\bf V}={\cal G}({\bf H}, {\bm \theta}_{\sf V})$, where ${\bm \theta}_{\sf V}$ includes all the trainable parameters in the GNN.
By using the theorems in \cite{hartford2018deep}, we can prove that ${\bf \Pi}_{\sf AN}^{\sf T}\hat{\bf V}{\bf \Pi}_{\sf UE}={\cal G}({\bf \Pi}_{\sf AN}^{\sf T}{\bf H}{\bf \Pi}_{\sf UE}, {\bm \theta}_{\sf V})$, which matches the PE property of the precoding policy.

The GNN that uses \eqref{eq: edge gnn upd1} as the update equation with sum-pooling function (i.e., ${\sf PL}_{k=1}^K(\cdot)=\sum_{k=1}^K(\cdot)$), mean-pooling function (i.e., ${\sf PL}_{k=1}^K(\cdot)=\frac{1}{K}\sum_{k=1}^K(\cdot)$) or max-pooling function  (i.e., ${\sf PL}_{k=1}^K(\cdot)=\max_{k=1}^K(\cdot)$) is referred to as \emph{Vanilla-GNN}.

The following proposition shows that the \emph{Vanilla-GNN} can only learn an element-wise function of ${\bf H}$ where  $v_{nk}$ only depends on $h_{nk}$,  when the problem scale is large.

\vspace{-2mm}\begin{proposition}\label{prop: GNN element wise policy}
	When $N\to\infty$ and $K\to\infty$, $h_{nk}$ is independently and identically distributed (i.i.d.) for $\forall n,k$, the Vanilla-GNN can only learn an element-wise precoding policy.
	\begin{IEEEproof}
		See Appendix \ref{proof: prop: GNN element wise policy}.
	\end{IEEEproof}
\end{proposition}

\vspace{-6mm}\begin{remark}
	\emph{
		As long as $q(\cdot)$ is a linear function, Proposition \ref{prop: GNN element wise policy} still holds for arbitrary non-linear combination function ${\sf CB}(\cdot)$ (e.g., a FNN as in \cite{GNN_mUE_RIS_JSAC_2021_Yu}).
	}
\end{remark}\vspace{-2mm}

The Vanilla-GNN can learn MRT well, which is an element-wise function of $\bf H$.
However, most precoding policies (e.g., ZFBF policy that is a mapping from the channel matrix to the pseudo-inverse of the channel matrix) are not element-wise functions of $\bf H$.
As a result, the training complexity of the Vanilla-GNN for learning non-element-wise policies is high and grows with $K$ and $N$ when $K$ and $N$ are finite. For example, it was shown in \cite{ZBC_WCNC} that even for a small-scale problem where $N=8$ and $K=4$, the Vanilla-GNN requires 200,000 training samples to achieve good performance for the SE-maximal precoding.

\vspace{-2mm}\section{A Model-GNN for Learning Precoding Policies} \label{sec: model-gnn}\vspace{-1mm}
In this section, we propose an edge GNN to learn precoding policies by resorting to the model of Taylor's expansion of matrix pseudo-inverse, which is called Model-GNN.
We first show that ZFBF can be accurately approximated by using the Taylor's expansion iteratively.
To provide useful insight for designing the Model-GNN, we then show the connection between the iterative Taylor's expansion with the Vanilla-GNN and explain why the Vanilla-GNN is not generalizable to problem scales when learning the ZFBF policy.
Next, we design the Model-GNN for learning the precoding policies in MISO systems, and finally extend it to coordinated beamforming in multi-cell systems.

\vspace{-4mm}\subsection{Connection between Vanilla-GNN and Iterative Taylor's Expansion of Pseudo-inverse}\vspace{-1mm}
Denote the pseudo-inverse of $\bf H$ as ${\bf H}^+$, whose first-order Taylor's expansion at ${\bf H}_0$ is ${\bf H}^+\approx 2{\bf H}_0^+ - {\bf H}_0^+ {\bf H}^{\sf H} {\bf H}_0^+$ \cite{magnus2019matrix}.
The Taylor's expansion can be used as an approximation of ${\bf H}^+$, which is closer to ${\bf H}^+$ than ${\bf H}_0^+$.
To obtain a more accurate approximation, we can approximate ${\bf H}^+$ with the first-order Taylor's expansion in an iterative manner, where in each iteration (say the $\ell$-th iteration), the pseudo-inverse is expanded at the approximation in the previous iteration, i.e.,\vspace{-8mm}
\begin{equation}\label{eq: pseudo inv upd}
{\bf D}^{(\ell)}=2{\bf D}^{(\ell-1)} - {\bf D}^{(\ell-1)} {\bf H}^{\sf H} {\bf D}^{(\ell-1)},\vspace{-2mm}
\end{equation}
where ${\bf D}^{(\ell)}=[d_{ij}^{(\ell)}]^{N\times K}$ is an approximation of ${\bf H}^+$, and ${\bf D}^{(0)}={\bf H}$ is the initial value.

As proved in Appendix \ref{proof: prop: contractive mapping}, \emph{${\bf D}^{(\ell)}$ can approximate ${\bf H}^+$ more accurately with more iterations}.

Denote ${\bf A}^{(\ell-1)}\triangleq {\bf H}^{\sf H} {\bf D}^{(\ell-1)}= [a_{ij}^{(\ell-1)}]^{K\times K}$, where $a_{ij}^{(\ell-1)}={\bf h}_i^{\sf H}\cdot{\bf d}_j^{(\ell-1)}$, ${\bf h}_i$ is the channel vector of UE$_i$, and ${\bf d}_j^{(\ell-1)}=[{ d}_{1j}^{(\ell-1)},\cdots, {d}_{Nj}^{(\ell-1)}]^{\sf T}$. Then, \eqref{eq: pseudo inv upd} can be re-written as ${\bf D}^{(\ell)}=2{\bf D}^{(\ell-1)} - {\bf D}^{(\ell-1)} {\bf A}^{(\ell-1)}$. By taking the element in the $n$th row and $k$th column of ${\bf D}^{(\ell)}$, we have \vspace{-1mm}
\begin{equation}\label{eq: iter upd element}
d_{nk}^{(\ell)} = 2d_{nk}^{(\ell-1)} - \underbrace{\textstyle\sum_{j=1}^K a_{jk}^{(\ell-1)} d_{nj}^{(\ell-1)}}_{(a)}.\vspace{-2mm}
\end{equation}

Next, we show the connection between the iteration equation with the update equation of the Vanilla-GNN. To this end, we provide the update equation of the Vanilla-GNN by taking summation as an example pooling function. Then,  \eqref{eq: edge gnn upd1} becomes\vspace{-2mm}
\begin{equation}\label{eq: edge gnn upd sum}
{\bf d}_{nk}^{(\ell)} = \sigma\Big({\bf S}{\bf d}_{nk}^{(\ell-1)} + {\bf P}\textstyle\sum_{i=1,i\neq n}^N{\bf d}_{ik}^{(\ell-1)} + {\bf Q}\textstyle\sum_{j=1,j\neq k}^K{\bf d}_{nj}^{(\ell-1)} \Big),\vspace{-2mm}
\end{equation}
which is the output vector of edge $(n,k)$ in the $\ell$th layer with $J_{\ell}$ elements, where $J_0=J_L=2$ and the values of $J_{\ell}$ for the hidden layers are tunable hyper-parameters. An example of updating ${\bf d}_{nk}^{(\ell)}$ with \eqref{eq: edge gnn upd sum} is given in Fig. \ref{fig-precoding-graph}(a).

The iteration equation in \eqref{eq: iter upd element} can be regarded as an update equation of a well-trained edge-GNN for learning the ZFBF policy (referred to as TGNN), where $d_{nk}^{(\ell)}$ is the output of edge $(n,k)$ in the $\ell$th layer, which is a scalar (i.e., $J_{\ell}=1$) instead of a vector as in \eqref{eq: edge gnn upd sum}. It is obtained by first aggregating the outputs of the edges connected to AN$_n$ in the $(\ell-1)$th layer (i.e., term $(a)$) and then combining with the output of edge $(n,k)$ in the $(\ell-1)$th layer. The aggregated information of the edges connected to UE$_k$ (i.e., the second term  in \eqref{eq: edge gnn upd sum}) is not in  \eqref{eq: iter upd element}, because this information has  been included in ${\bf A}^{(\ell-1)}$. Specifically, the $k$th column of ${\bf A}^{(\ell-1)}$, ${\bf a}_k^{(\ell-1)}=\sum_{n=1}^N \vec{\bf h}_n d_{nk}^{(\ell-1)}$, is the aggregated information of the edges connected to UE$_k$, where $\vec{\bf h}_n=[h_{n1},\cdots,h_{nK}]^{\sf T}$ is the $n$th row of ${\bf H}$. It is shown from \eqref{eq: iter upd element}  that the weights on the output of the edges (i.e., $a_{jk}^{(\ell-1)}$) are different in the aggregator, rather than being identical (i.e., ${\bf Q}$) as in \eqref{eq: edge gnn upd sum}. Hence, both the combination and processing functions in the TGNN differ from the Vanilla-GNN.

From  the iteration equation in \eqref{eq: iter upd element}, the outputs of all the edges connected to UE$_k$ in the well-trained TGNN 
are updated as\vspace{-2mm}
\begin{equation}\label{eq: iter upd vec}
{\bf d}_{k}^{(\ell)} = 2{\bf d}_{k}^{(\ell-1)} - \textstyle\sum_{j=1}^K a_{jk}^{(\ell-1)} {\bf d}_{j}^{(\ell-1)}=2{\bf d}_{k}^{(\ell-1)} - \textstyle\sum_{j=1}^K ({\bf h}_j^{\sf H}{\bf d}_k^{(\ell-1)}) {\bf d}_{j}^{(\ell-1)}.\vspace{-1mm}
\end{equation}
When $\ell=0$, ${\bf d}_k^{(0)}={\bf h}_k$. Then, ${\bf d}_{k}^{(1)}=2{\bf h}_k - \sum_{j=1}^K ({\bf h}_j^{\sf H}{\bf h}_k) {\bf h}_{j}$, where the weight $a_{jk}^{(0)}={\bf h}_j^{\sf H}{\bf h}_k$ reflects the orthogonality of ${\bf h}_j$ and ${\bf h}_k$, which can be seen as finding the orthogonal complement of the space spanned with $\{{\bf h}_j, j=1,\cdots,K, j\neq k\}$.  If ${\bf h}_j$ tends to be orthogonal with ${\bf h}_k$, less component is subtracted from ${\bf h}_k$. The orthogonal complement can be found after iterations until ${\bf h}_j^{\sf H}\cdot{\bf d}_k^{(\ell)}\to 0$. When $\ell=L$, ${\bf d}_k^{(L)}=\hat {\bf v}_k$, which is the learned ZFBF vector for UE$_k$.


\vspace{-4mm}\subsection{Interpreting the Generalization Ability to Number of Users} \label{sec: dimension generalize}\vspace{-1mm}
In the update equation of the Vanilla-GNN,  the dimensions of weight matrices ${\bf S, P}$ and ${\bf Q}$ are independent of $K, N$ due  to the parameter sharing  \cite{ZBC_WCNC}. As a result, after the GNN is trained by the samples generated with $K$ users and $N$ antennas,
it can output a $K'\times N'$ precoding matrix using the same update equation  in the inference stage, where $K'\neq K$ and $N'\neq N$.
However, this does not mean that the Vanilla-GNN can perform well if there are $K'$ users and $N'$ antennas during inference.
In what follows, we use an example to demonstrate that the Vanilla-GNN can not be generalized to $K$. The generalization ability to $N$ can be explained similarly.

\textbf{Example}: For the MISO system where the BS is with $N$ antennas, we use the Vanilla-GNN to learn a ZFBF policy, considering small- and large-scale problems respectively with $K$ and $2K$ users. In the large-scale problem, the users are divided into two sets, denoted as ${\cal U}_1$ and ${\cal U}_2$, respectively. The channel matrices for users in ${\cal U}_1$ and ${\cal U}_2$ are respectively ${\bf H}_1$ and ${\bf H}_2$, which are orthogonal to each other, i.e., ${\bf H}_1^{\sf H}{\bf H}_2={\bf 0}$. Then, the ZFBF policy for ${\bf H}=[{\bf H}_{1}^{\sf T}, {\bf H}_{2}^{\sf T}]^{\sf T}$ is,
\begin{eqnarray}\label{eq: opt v}
{\bf V}_{\sf ZF}({\bf H})={\bf H}({\bf H}^{\sf H}{\bf H})^{-1}
=[{\bf H}_1, {\bf H}_2]\Bigg(\begin{bmatrix}{\bf H}_{1}^{\sf H}\\{\bf H}_{2}^{\sf H}\end{bmatrix}\cdot [{\bf H}_1, {\bf H}_2]\Bigg)^{-1}\!\!\!
=[\underbrace{{\bf H}_1({\bf H}_1^{\sf H}{\bf H}_1)^{-1}}_{{\bf V}_{\sf ZF}({\bf H}_1)},\underbrace{{\bf H}_2({\bf H}_2^{\sf H}{\bf H}_2)^{-1}}_{{\bf V}_{\sf ZF}({\bf H}_2)}],\vspace{-1mm} 
\end{eqnarray}
where the  beamforming vectors for the users in ${\cal U}_1$ and ${\cal U}_2$ are decoupled.

As proved in Appendix \ref{proof: prop: gnn scalability}, even if a well-trained Vanilla-GNN can accurately approximate the ZFBF policy for the small-scale problem (i.e., ${\cal G}({\bf H}_1, \bm{\theta}_{\sf V}^{\star})={\bf V}_{\sf ZF}({\bf H}_1)$, ${\cal G}({\bf H}_2, \bm{\theta}_{\sf V}^{\star})={\bf V}_{\sf ZF}({\bf H}_2)$), where $\bm{\theta}_{\sf V}^{\star}$ include the trained weight matrices, the Vanilla-GNN cannot learn ${\bf V}_{\sf ZF}({\bf H})$ for the large-scale problem with $2K$ users (i.e., ${\cal G}({\bf H}, \bm{\theta}_{\sf V}^{\star}) \neq {\bf V}_{\sf ZF}({\bf H})$).

\begin{figure}[!htb]
	\centering
	\includegraphics[width=0.9\linewidth]{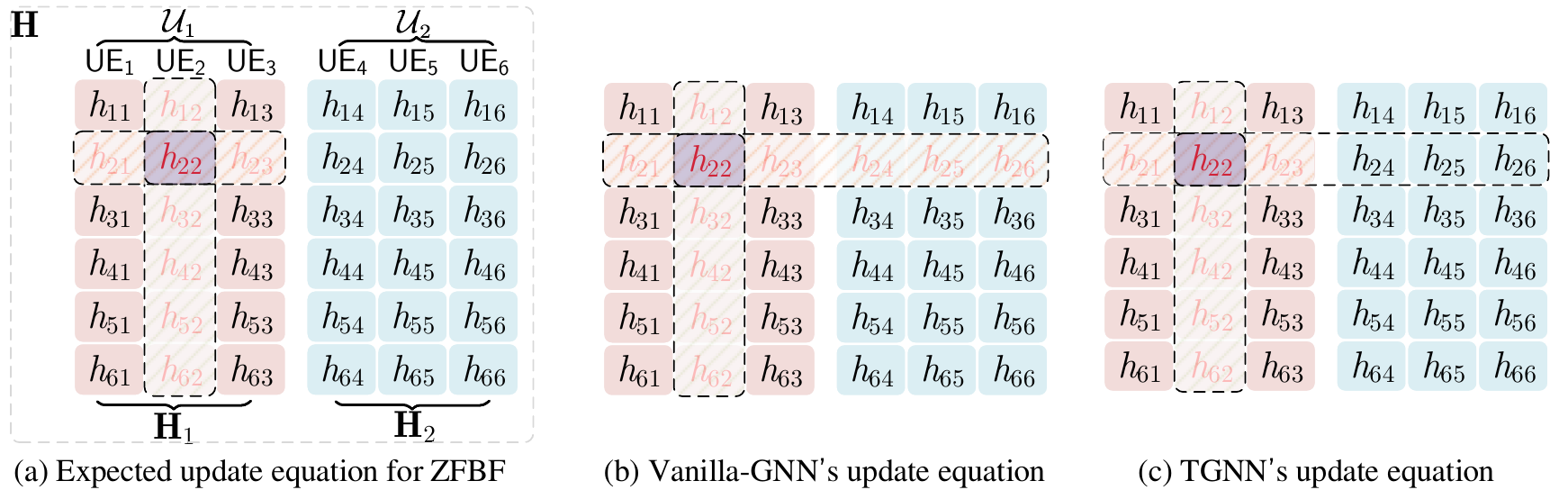}\vspace{-3mm}
	\caption{Example of the update equations for learning the ZFBF policy, $N=6, K=6$, ${\bf H}=[{\bf H}_{1}, {\bf H}_{2}]$, and ${\bf H}_1^{\sf H}{\bf H}_2={\bf 0}$.}
	\label{fig-update-example}
	\vspace{-2mm}
\end{figure}

To help understand, we consider a specific system setting as illustrated in Fig. \ref{fig-update-example}, where ${\cal U}_1=\{\text{UE}_1, \text{UE}_2, \text{UE}_3\}, {\cal U}_2=\{\text{UE}_4, \text{UE}_5, \text{UE}_6\}$. Since ${\bf H}_1^{\sf H}{\bf H}_2={\bf 0}$, the update equation for the ZFBF vectors of users in ${\cal U}_1$ and ${\cal U}_2$ are decoupled. Take the update for edge $(2, 2)$ as an example. When aggregating information from the edges neighbored to edge $(2, 2)$, only the elements in ${\bf H}_1$ should be aggregated, as shown in the dashed squares in Fig. \ref{fig-update-example}(a).

However, the information in ${\bf H}_2$ is also aggregated by the Vanilla-GNN because the GNN uses the same weight matrix ${\bf Q}$ to aggregate information from all the edges neighbored to edge $(2, 2)$ by AN$_2$, as shown in Fig. \ref{fig-update-example}(b).
Since the analysis is the same for all layers, the Vanilla-GNN cannot yield ZFBF matrix in the output layer.

By contrast, the well-trained TGNN can approximate the ZFBF policy accurately in both the small- and large-scale problems according to the proof in Appendix \ref{proof: prop: contractive mapping}, despite that it also aggregates information from all edges in each layer (shown in Fig. \ref{fig-update-example}(c)) as the Vanilla-GNN. This is because the information of the edges neighbored to edge $(2, 2)$ is assigned with different weights during aggregation, which are zero when aggregating information in ${\bf H}_2$.

One may wonder whether or not a Vanilla-GNN can be generalized to $K$ if FNNs are used as the combination and processing functions as in \cite{GNN_mUE_RIS_JSAC_2021_Yu} (this GNN is referred to as \emph{Vanilla-GNN-FC}). To answer this question, we use the Vanilla-GNN-FC to learn the ZFBF policy by approximating the update equation of the well-trained TGNN in \eqref{eq: iter upd vec} as,\vspace{-1mm}
\begin{equation}\label{eq: vanilla-gnn-fc}
{\bf d}_{k}^{(\ell)} = {\sf CB}\big([{\bf h}_{k}, {\bf d}_{k}^{(\ell-1)}], \textstyle\sum_{j=1,j\neq k}^K q([{\bf h}_j, {\bf d}_{j}^{(\ell-1)}])\big),\vspace{-1mm}
\end{equation}
where ${\sf CB}(\cdot)$ and $q(\cdot)$ are FNNs and the trainable parameters are omitted for notational simplicity. Parameter sharing can be introduced into the FNNs to guarantee that the ${\bf D}^{(\ell)}$ is equivariant to the permutation of antennas (i.e., rows of ${\bf D}^{(\ell-1)}$ and ${\bf H}$). 

By comparing \eqref{eq: vanilla-gnn-fc} with \eqref{eq: iter upd vec}, we can see  that ${\bf d}_{k}^{(\ell-1)}$ in ${\sf CB}(\cdot)$ is used to compute the weight  on ${\bf d}_{j}^{(\ell-1)}$, i.e., $a_{jk}^{(\ell-1)}={\bf h}_j^{\sf H}{\bf d}_k^{(\ell-1)}$. This indicates that after aggregation (i.e., after passing through $q(\cdot)$ and the sum-pooling function), the elements in ${\bf H}_{/k}\triangleq[{\bf h}_1,\cdots,{\bf h}_{k-1}, {\bf h}_{k+1},\cdots,{\bf h}_K]^{N\times (K-1)}$ and ${\bf D}_{/k}^{(\ell-1)}\triangleq[{\bf d}_1^{(\ell-1)},\cdots,{\bf d}_{k-1}^{(\ell-1)}, {\bf d}_{k+1}^{(\ell-1)},\cdots,{\bf d}_K^{(\ell-1)}]^{N\times (K-1)}$ should be recoverable, such that the inner product of ${\bf d}_{k}^{(\ell-1)}$ and each vector in ${\bf H}_{/k}$ can be computed to obtain the weight on each vector in ${\bf D}_{/k}^{(\ell-1)}$. This requires the  existence of a one-to-one mapping ${\rho}(\cdot):\sum_{j=1,j\neq k}^K q([{\bf h}_j, {\bf d}_{j}^{(\ell-1)}])\mapsto [{\bf H}_{/k}^{\sf T}, {\bf D}_{/k}^{(\ell-1){\sf T}}]^{\sf T}$. According to \cite{zaheer2017deep}, when the output dimension of $q(\cdot)$ is no less than the dimension of $[{\bf H}_{/k}^{\sf T}, {\bf D}_{/k}^{(\ell-1){\sf T}}]^{\sf T}$ (i.e., $2N\times (K-1)$), the one-to-one mapping exists. Then, ${\sf CB}(\cdot)$ can be expressed as a composite function of $\rho(\cdot)$ (denoted as $\phi(\cdot)$) as,\vspace{-1mm}
\begin{eqnarray}\label{eq: vanilla-gnn-fc 1}
&&{\sf CB}\big([{\bf h}_{k}, {\bf d}_{k}^{(\ell-1)}], \textstyle\sum_{j=1,j\neq k}^K q([{\bf h}_j, {\bf d}_{j}^{(\ell-1)}])\big)
= \phi\big([{\bf h}_{k}, {\bf d}_{k}^{(\ell-1)}], \rho(\textstyle\sum_{j=1,j\neq k}^K q([{\bf h}_j, {\bf d}_{j}^{(\ell-1)}]))\big)\notag\\
&=&\phi([{\bf h}_{k}, {\bf d}_{k}^{(\ell-1)}], [{\bf H}_{/k}^{\sf T}, {\bf D}_{/k}^{(\ell-1){\sf T}}]^{\sf T})
\triangleq 2{\bf d}_{k}^{(\ell-1)} - ({\bf h}_k^{\sf H}{\bf d}_k^{(\ell-1)}) {\bf d}_{k}^{(\ell-1)} - {\bf D}_{/k}^{(\ell-1)}({\bf H}_{/k}^{\sf H}{\bf d}_{k}^{(\ell-1)})\notag\\
&=& 2{\bf d}_{k}^{(\ell-1)} - \textstyle\sum_{j=1}^K ({\bf h}_j^{\sf T}{\bf d}_k^{(\ell-1)}) {\bf d}_{j}^{(\ell-1)}.\vspace{-1mm}
\end{eqnarray}
This indicates that if the output dimension of $q(\cdot)$ exceeds $2N\times (K-1)$ and two FNNs can well-approximate the processing and combination functions, the Vanilla-FNN-FC trained with the samples of $K$ users can learn the ZFBF policy for $K$ users with good performance.

When the trained Vanilla-GNN-FC is tested in a system with $K'>K$ users, $[{\bf H}_{/k}^{\sf T}, {\bf D}_{/k}^{(\ell-1){\sf T}}]^{\sf T}\in{\mathbb C}^{2N\times (K'-1)}$, whose dimension is higher than the output dimension of $q(\cdot)$ in the trained Vanilla-GNN-FC. Then, $[{\bf H}_{/k}^{\sf T}, {\bf D}_{/k}^{(\ell-1){\sf T}}]^{\sf T}$ is ``compressed'' by the aggregation, and hence cannot be recovered from $\sum_{j=1,j\neq k}^{K'} q([{\bf h}_j, {\bf d}_{j}^{(\ell-1)}])$ with $\rho(\cdot)$. As a consequence, the different weights used to aggregate the output of the edges (i.e., $a_{jk}^{(\ell-1)}$) cannot be obtained. This indicates that the generalizable ability of a Vanilla-GNN to the scales of precoding problems cannot be improved by simply using FNNs as the combination and processing functions.

\vspace{-3mm}
\subsection{Structure of the Proposed Model-GNN} \label{sec: model-gnn-complexiy}\vspace{-1mm}
For notational simplicity, we consider the case that the feature and output of each edge in the GNN is a real-valued scalar in this subsection (i.e., $J_\ell=1, \forall \ell$), unless otherwise specified.

The iteration equation in \eqref{eq: iter upd element} can only be used to approximate matrix pseudo-inverse. As analyzed previously by the example, the different weights for aggregating the information of the edges are critical for a GNN to learn the ZFBF policy with generalizability to $K$. To learn the precoding policies that consist of pseudo-inverse operation explicitly as in \eqref{eq: opt precod struc} or implicitly, we introduce the two terms in \eqref{eq: iter upd element} into the update equation  in \eqref{eq: edge gnn upd sum}.
Specifically, in the $\ell$th layer, the output of edge $(n,k)$ is obtained by the following two steps.

\textbf{Step 1}: The two terms in \eqref{eq: iter upd element} are concatenated as a vector ${\bf c}_{nk}^{(\ell-1)}=[{d}_{nk}^{(\ell-1)}, \sum_{j=1}^K {a}_{jk}^{(\ell-1)}{d}_{nj}^{(\ell-1)} ]^{\sf T}$. Then, the two terms can be weighted with trainable parameters in the next step (instead of  weighted with constants ``2'' and ``-1" in \eqref{eq: iter upd element}) for learning the power and direction in  precoding.

\textbf{Step 2}: The output of edge $(n,k)$ is obtained by replacing  ${\bf d}_{nk}^{(\ell-1)}$ by ${\bf c}_{nk}^{(\ell-1)}$  in \eqref{eq: edge gnn upd sum}, i.e.,\vspace{-1mm}
\begin{eqnarray}\label{eq: upd propose GNN 1}
{d}_{nk}^{(\ell)} &=& \sigma\Big({\bf s}^{'\sf T}{\bf c}_{nk}^{(\ell-1)} + {\bf p}^{\sf T}\cdot\textstyle\sum_{i=1,i\neq n}^N {\bf c}_{ik}^{(\ell-1)} + {\bf q}^{\sf T}\cdot \textstyle\sum_{j=1,j\neq k}^K{\bf c}_{nj}^{(\ell-1)}\Big) \notag\\
&=& \sigma\Big({\bf s}^{\sf T}{\bf c}_{nk}^{(\ell-1)} + {\bf p}^{\sf T}\cdot\textstyle\sum_{i=1}^N {\bf c}_{ik}^{(\ell-1)} + {\bf q}^{\sf T}\cdot \textstyle\sum_{j=1}^K{\bf c}_{nj}^{(\ell-1)}\Big),\vspace{-1mm}
\end{eqnarray}
where ${\bf s}\triangleq{\bf s}'-{\bf p}-{\bf q}$, ${\bf s}'$, ${\bf p}$ and ${\bf q}$ are trainable parameters, which are vectors instead of matrices as in \eqref{eq: edge gnn upd sum} since ${d}_{nk}^{(\ell)}$ is a scalar. The superscript $(\ell)$ of all the weight vectors are again omitted.

\begin{figure}[!htb]
	\centering
	\includegraphics[width=0.65\linewidth]{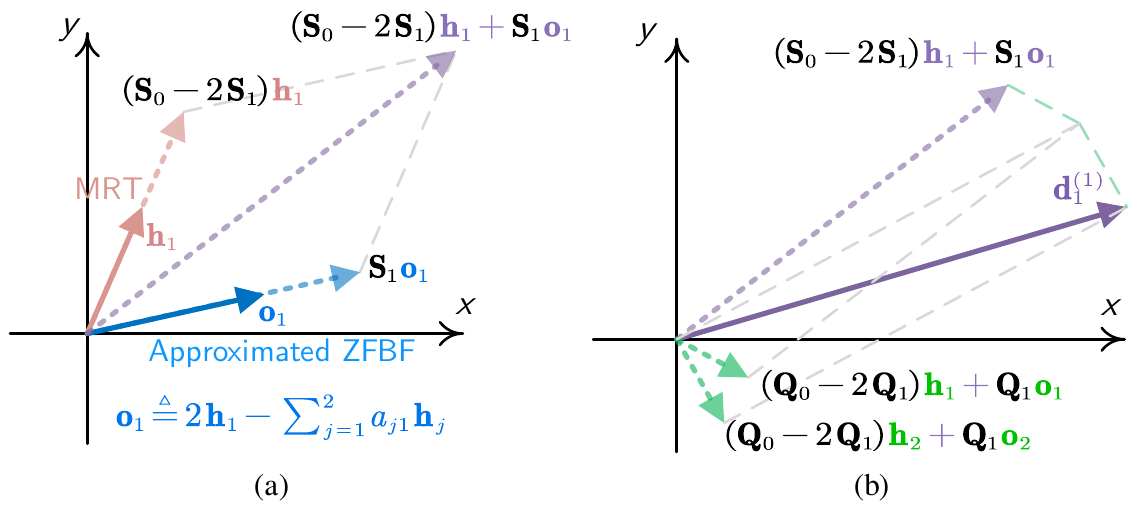}\vspace{-3mm}
	\caption{An example of updating ${\bf d}_1^{(1)}$ for UE$_1$, $K=2, N=2$. The superscript $(1)$ in the vectors is omitted. In (a), ${\bf p}={\bf q}={\bf 0}$, by training ${\bf S}_0$ and ${\bf S}_1$, ${\bf d}_1^{(1)}$ can be adjusted to the purple dashed vector. In (b), ${\bf q}\neq {\bf 0}$, by aggregating the adjusted precoding vector of UE$_1$ and UE$_2$ (i.e., the green dashed vectors), ${\bf d}_1^{(1)}$ can be adjusted to the purple solid vector.}\label{fig-examp-learnw} 
	
\end{figure}

To understand how a precoding policy is learned by the update equation in \eqref{eq: upd propose GNN 1}, we first analyze the role of weight vector ${\bf s}$ by setting ${\bf p}={\bf q}={\bf 0}$, and omit the activation function. In this case, \eqref{eq: upd propose GNN 1} becomes $d_{nk}^{(\ell)}={\bf s}^{\sf T}{\bf c}_{nk}^{(\ell-1)}={s}_0d_{nk}^{(\ell-1)} + {s}_1\sum_{j=1}^K {a}_{jk}^{(\ell-1)}{d}_{nj}^{(\ell-1)}$, where ${\bf s}=[s_0, s_1]^{\sf T}$, and ${\bf d}_{k}^{(\ell)}=[d_{1k}^{(\ell)},\cdots,d_{Nk}^{(\ell)}]^{\sf T}$ is the output vector for UE$_k$ in the $\ell$th layer (${\bf d}_{k}^{(L)}$ is the precoding vector for UE$_k$). When $\ell=0$, ${\bf d}_{k}^{(0)}={\bf h}_k$, and ${\bf d}_{k}^{(1)}$ can be re-written as,
\begin{equation}\label{eq: upd model-gnn simple}
{\bf d}_{k}^{(1)}={\bf S}_0{\bf d}_{k}^{(0)} + {\bf S}_1\textstyle\sum_{j=1}^K  {a}_{jk}^{(\ell-1)}{\bf d}_{j}^{(0)}
= \underbrace{({\bf S}_0-2{\bf S}_1){\bf h}_{k}}_{(a)} + \underbrace{{\bf S}_1 \big(2{\bf h}_{k}-\textstyle\sum_{j=1}^K  {a}_{jk}^{(0)}{\bf h}_{j}\big)}_{(b)},\vspace{-1mm}
\end{equation}
where ${\bf S}_0$ and ${\bf S}_1$ are diagonal matrices with diagonal elements being ${s}_0$ and ${s}_1$, respectively. The terms $(a)$ and $(b)$ in \eqref{eq: upd model-gnn simple} are respectively the scaled version of the MRT vector and the approximated ZFBF vector with Taylor's expansion for UE$_k$. By learning ${s}_0$ and ${s}_1$, both the direction and power of ${\bf d}_{k}^{(1)}$ can be adjusted, as shown in Fig. \ref{fig-examp-learnw}(a).

The role of ${\bf p}$ and ${\bf q}$ are to aggregate information from the edges neighbored to edge $(n,k)$ by UE$_k$ and AN$_n$, respectively.
We take the role of ${\bf q}=[q_0, q_1]^{\sf T}$ as an example by setting ${\bf p}=[p_1, p_2]={\bf 0}$. By using the same way as we  obtain \eqref{eq: upd model-gnn simple}, ${\bf d}_k^{(1)}$ can be re-written as,\vspace{-1mm}
\begin{equation}\label{eq: upd model-gnn 1}
{\bf d}_{k}^{(1)}=({\bf S}_0-2{\bf S}_1){\bf h}_{k} + {\bf S}_1 \big(2{\bf h}_{k}-\sum_{j=1}^K  {a}_{jk}^{(0)}{\bf h}_{j}\big) + \sum_{j=1}^K \Big(({\bf Q}_0-2{\bf Q}_1){\bf h}_{j} + {\bf Q}_1 \big(2{\bf h}_{j}-\sum_{j=1}^K  {a}_{ij}^{(0)}{\bf h}_{i}\big)\Big),\vspace{-1mm}
\end{equation}
where ${\bf Q}_0$ and ${\bf Q}_1$ are diagonal matrices with diagonal elements as $q_0$ and $q_1$, respectively. Compared with \eqref{eq: upd model-gnn simple}, an extra term  is added  in \eqref{eq: upd model-gnn 1} to update ${\bf d}_k^{(1)}$ (as shown in the green vectors in Fig. \ref{fig-examp-learnw} (b) for $k=1$), which is the summation of the scaled version of the MRT vectors and the scaled version of the
approximated ZFBF vectors for all the $K$ users. The two weight matrices ${\bf Q}_0-2{\bf Q}_1$ and ${\bf Q}_1$ are identical among users, which do not depend on the orthogonality of channel vectors.
Since the precoding vectors for users are still coupled even if the users are with orthogonal channels due to the power constraint in problem \emph{P1}, ${\bf Q}_0$ and ${\bf Q}_1$ capture the dependence among the precoding vectors during the update procedure.

In what follows, we introduce the update equation for Model-GNN. After substituting ${\bf c}_{nk}^{(\ell-1)}=[{d}_{nk}^{(\ell-1)}, \sum_{j=1}^K {a}_{jk}^{(\ell-1)}{d}_{nj}^{(\ell-1)} ]^{\sf T}$, ${\bf s}=[s_0, s_1]^{\sf T}$, ${\bf p}=[p_0, p_1]^{\sf T}$, and ${\bf q}=[q_0, q_1]^{\sf T}$, \eqref{eq: upd propose GNN 1} can be further re-written as,
\begin{eqnarray}\label{eq: upd propose GNN}
{d}_{nk}^{(\ell)} \!\!&\!\!=\!\!&\!\!
\sigma\Big(s_0 {d}_{nk}^{(\ell-1)}
+ s_1 \textstyle\sum_{j=1}^K {a}_{jk}^{(\ell-1)}{d}_{nj}^{(\ell-1)}
+ p_0\cdot\sum_{i=1}^N {d}_{ik}^{(\ell-1)} + \notag\\
&& ~~~~p_1 \cdot\textstyle\sum_{i=1}^N \sum_{j=1}^K {a}_{jk}^{(\ell-1)} {d}_{ij}^{(\ell-1)}
+ q_0\cdot \sum_{j=1}^K {d}_{nj}^{(\ell-1)}
+ q_1 \cdot \sum_{j=1}^K \sum_{m=1}^K {a}_{jm}^{(\ell-1)} {d}_{nj}^{(\ell-1)} \Big) \notag\\
\!\!&\!\!=\!\!&\!\! \sigma\Big(s_0 {d}_{nk}^{(\ell-1)}
+ \underbrace{\textstyle\sum_{i=1}^N p_0 {d}_{ik}^{(\ell-1)}}_{(a)}
+ \underbrace{\textstyle\sum_{j=1}^K \bar{q}_{jk} d_{nj}^{(\ell-1)}}_{(b)}
+ \underbrace{\textstyle\sum_{i=1}^N \sum_{j=1}^K p_1{a}_{jk}^{(\ell-1)} {d}_{ij}^{(\ell-1)} }_{(c)} \Big),
\end{eqnarray}
where $\bar{q}_{jk}\triangleq\textstyle\big(q_0 + s_1{a}_{jk}^{(\ell-1)} + q_1 \sum_{m=1}^K {a}_{jm}^{(\ell-1)}\big)$.

In \eqref{eq: upd propose GNN}, term $(a)$ aggregates the outputs of all the edges connected to UE$_k$ with same weight $p_0$. Term $(b)$ aggregates the outputs of all the edges connected to AN$_n$ (i.e., the interfering links) with different weights $\bar{q}_{jk}$.  Term $(c)$ aggregates the outputs from all the edges that are not neighbored to edge $(n,k)$, since the outputs of neighbored edges are aggregated twice, i.e., when computing the second term in ${\bf c}_{nk}^{(\ell-1)}$ in step 1 and when passing through a GNN layer in step 2. The complexity of computing term $(c)$ is higher than computing term $(a)$ and $(b)$ because term $(c)$ aggregates more information than the other two terms.
To reduce the computational complexity  without losing the expression ability of the GNN, we can omit the term $(c)$, because the information of non-neighbored edges can be aggregated after stacking multiple layers.

We refer to the GNN with the updating equation in \eqref{eq: upd propose GNN} as \emph{Model-GNN}.
The output of edge $(n,k)$ in each layer (say the $\ell$th layer) of Model-GNN can be a vector, i.e., ${\bf d}_{nk}^{(\ell)}$, which contains $J_{\ell}>1$ elements. In the input and output layer of Model-GNN, ${\bf d}_{nk}^{(0)}=[{\sf Re}(h_{nk}), {\sf Im}(h_{nk})]$, and ${\bf d}_{nk}^{(L)}=[{\sf Re}(\hat{v}_{nk}), {\sf Im}(\hat{v}_{nk})]$, i.e., $J_0=J_L=2$. Analogous to the Vanilla-GNN,
when $0<\ell<L$, $J_{\ell}$ can exceed two to improve the expression ability of the Model-GNN,
and then the trainable parameters $s_0, s_1, p_0, p_1, q_0, q_1$ become matrices instead of scalars.
An example of updating ${\bf d}_{nk}^{(\ell)}$ with Model-GNN is given in Fig. \ref{fig-precoding-graph}(b).
The input-output relationship of the Model-GNN is denoted as $\hat{\bf V}={\cal G}_{\sf M}({\bf H},{\bm \theta}_{\sf M})$, where ${\bm \theta}_{\sf M}$ includes all the trainable parameters.
It is not hard to prove that $\hat{\bf V}={\cal G}_{\sf M}({\bf H}, {\bm \theta}_{\sf M})$ is 2D-PE to $\bf H$.

The Model-GNN can be trained in an unsupervised manner, hence labels are not required by solving optimization problems. For problem \emph{P1}, we train the Model-GNN to minimize the negative SE averaged over all the training samples, and the power constraint can be satisfied by passing the output of the GNN $\hat{\bf V}$ through a power normalization layer, i.e., $P_{\max}\frac{\hat{\bf V}}{\|\hat{\bf V}\|}$.

\vspace{-5mm}\subsection{Computation Complexities} \label{sec: cpt cplxty} \vspace{-2mm}
For a $N\times K$ matrix ${\bf A}$ and a $K\times M$ matrix ${\bf B}$, $MNK$ additions and multiplications are required for computing ${\bf A}{\bf B}$, hence the number of floating-point operations (FLOPs) for the matrix multiplication is $2MNK$.
Recall that ${\bf d}_{nk}^{(\ell)}$ in \eqref{eq: edge gnn upd sum} contains $J_{\ell}$ elements, each weight matrix is with size of $J_{\ell+1}\times J_{\ell}$, and ${\bf D}^{(\ell)}=[{\bf d}_{nk}^{(\ell)}]^{N\times K}$.

For the Vanilla-GNN, $2J_{\ell+1}J_{\ell}$ operations are required to compute  each of the three terms in \eqref{eq: edge gnn upd sum}. Since matrix multiplication needs to be computed for each of the $NK$ elements in ${\bf D}^{(\ell)}$, the total number of operations is $6NKJ_{\ell+1}J_{\ell}$. The second and third terms in \eqref{eq: edge gnn upd sum} are respectively the summation of the $k$th column and $n$th row of ${\bf D}^{(\ell)}$, which requires $NJ_{\ell}$ and $KJ_{\ell}$ addition operations for computing. Since there are $K$ columns and $N$ rows in ${\bf D}^{(\ell)}$, the total number of addition operations is $KNJ_{\ell}+NKJ_{\ell}=2NKJ_{\ell}$. Hence, the total number of FLOPs of the Vanilla-GNN with $(L+1)$ layers is $\sum_{\ell=0}^{L-1}(6NKJ_{\ell+1}J_{\ell}+2NKJ_{\ell})$.

For the Model-GNN, the update equation is obtained by passing ${\bf D}^{(\ell-1)}$ through the two steps above and then omitting term $(c)$ in \eqref{eq: upd propose GNN}. In step 1, we need to compute $\sum_{j=1}^K {a}_{jk}^{(\ell-1)}{d}_{nj}^{(\ell-1)}$, which is the element in the $n$th row and $k$th column of ${\bf D}^{(\ell-1)}{\bf A}^{(\ell-1)}={\bf D}^{(\ell-1)}{\bf H}^{\sf H}{\bf D}^{(\ell-1)}$. Since it requires $2NK^2$ operations to respectively compute ${\bf A}^{(\ell-1)}={\bf H}^{\sf H}{\bf D}^{(\ell-1)}$ and ${\bf D}^{(\ell-1)}{\bf A}^{(\ell-1)}$, the total number of FLOPs for computing ${\bf D}^{(\ell-1)}{\bf H}^{\sf H}{\bf D}^{(\ell-1)}$ is $4NK^2$, which is multiplied by $J_{\ell}$ when $J_{\ell}>1$. In step 2, ${\bf c}_{nk}^{(\ell-1)}$ with $2J_{\ell}$ elements is passed through a Vanilla-GNN layer, which requires $12NKJ_{\ell+1}J_{\ell}+4NKJ_{\ell}$ operations. 
In term $(c)$, computing $\sum_{i=1}^N \sum_{j=1}^K {a}_{jk}^{(\ell-1)} {d}_{ij}^{(\ell-1)}$ requires $2NK$ operations and multiplying $p_1$ requires additional one operation, which are respectively multiplied by $J_{\ell}$ and $J_{\ell+1}J_{\ell}$ when $J_{\ell}, J_{\ell+1}>1$. Since term $(c)$ does not change with $n$ and only change with $k$, the number of FLOPs reduced by omitting term $(c)$ in each layer is $2NK^2J_{\ell} + 2KJ_{\ell+1}J_{\ell}$.
Hence, the total number of FLOPs of Model-GNN with $(L+1)$ layers is $\sum_{\ell=0}^{L-1}(2NK^2J_{\ell}+12NKJ_{\ell+1}J_{\ell}-2KJ_{\ell+1}J_{\ell}+4NKJ_{\ell})$.

\vspace{-3mm}
\subsection{Extension to Multi-cell Networks}\label{subsec: multicell} \vspace{-1mm}
The Model-GNN can also be used to learn coordinate beamforming policy in $M$ cells \cite{CoMP_WCM_2013}. Denote ${\bf V}_m=[{\bf v}_{1_m},\cdots,{\bf v}_{K_m}]\in{\mathbb C}^{N\times K}$ as the precoding matrix of the $m$th BS, where ${\bf v}_{k_m}$ is the precoding vector of the $k$th UE in the $m$th cell (denoted as UE$_{k_m}$), $k=1,\cdots, K, m=1,\cdots, M$. We take the SE-maximal coordinate beamforming as an example, i.e.,
\begin{subequations}\label{P1: SE max-multicell}
	\begin{align}
	\emph{P2:}~\max_{{\bf V}_1,\cdots,{\bf V}_M} \sum_{m=1}^M\sum_{k=1}^K \log_2\Bigg(1 + \frac{|{\bf h}_{k_m,m}^{\sf H}{\bf v}_{k_m}|^2}{\sum_{(j, i)\neq(k, m)}|{\bf h}_{k_m,i}^{\sf H}{\bf v}_{j_i}|^2 + \sigma_0^2}\Bigg), ~~
	{\rm s.t.}~ {\sf Tr}({\bf V}_m^{\sf H}{\bf V}_m) \leq P_{\max}, \forall m,\notag \label{P1-m: constraint}
	\end{align}
\end{subequations}
where ${\bf h}_{k_{m},i}$ is the channel vector from the antennas at the $i$th BS to UE$_{k_m}$.

The precoding graph contains $MN$ antenna vertexes, $MK$ user vertexes, and edges.  We call the links between the antennas and users in a single cell as \emph{intra-cell edges}, and the links across cells (which generates inter-cell interference) as \emph{inter-cell edges}.
The update process of the Model-GNN in the multi-cell system is as follows, which differs from the single-cell case in computing the  inner product and sharing the trainable parameters.
We take the update of output of an intra-cell edge $(n_m, k_m)$ (i.e., the edge connecting AN$_n$ and UE$_k$ in the $m$th cell) as an example, and the update of outputs of inter-cell edges are similar.
We again consider that the output of each edge is a real-valued scalar for notational simplicity.

\textbf{Step 1}: The intermediate output of edge $(n_m, k_m)$ is obtained by ${\bf c}_{n_m k_m}^{(\ell-1)}=[d_{n_m k_m}^{(\ell-1)}, \sum_{m'=1}^M\sum_{j=1}^K a_{j_{m'} k_m}^{(\ell-1)} d_{n_m j_{m'}}^{(\ell-1)}]^{\sf T}$, where $a_{j_{m'} k_m}^{(\ell)}= {\bf h}_{j_{m'},m}^{\sf H}{\bf d}_{m,k_{m}}^{(\ell-1)}$, ${\bf d}_{m,j_{m'}}^{(\ell)}=[{d}_{1_m j_{m'}}^{(\ell)},\cdots, {d}_{N_m j_{m'}}^{(\ell)}]^{\sf T}$ is the output of the edges connected with UE$_{j_{m'}}$ and the antennas at the $m$th BS. Different from computing ${\bf c}_{nk}^{(\ell-1)}$ in the single-cell case,  only the information of edges connecting antennas in the $m$th cell (instead of all the edges) are considered when computing the  inner product $a_{j_{m'} k_m}^{(\ell)}= {\bf h}_{j_{m'},m}^{\sf H}{\bf d}_{m,k_{m}}^{(\ell-1)}$.

\textbf{Step 2}:
The coordinate beamforming policy is equivariant to the permutation of cells, permutations of the antennas at each BS and permutations of the users in each cell,  which no longer satisfies the same PE property as the single-cell precoding. With the method to design parameter sharing for satisfying a given PE property in  \cite{GJ_TWC_GNN}, the output of edge $(n_m, k_m)$ is,
\begin{equation}
d_{n_m k_m}^{(\ell)} = \sigma\Bigg({\bf s}^{\sf T}{\bf c}_{n_m k_m}^{(\ell-1)}+{\bf p}_{\sf a}^{\sf T}\sum_{\substack{j=1\\j\neq n}}^N{\bf c}_{j_m k_m}^{(\ell-1)} + {\bf p}_{\sf r}^{\sf T}\sum_{\substack{m'=1\\m'\neq m}}^M\sum_{\substack{j=1\\j\neq n}}^N{\bf c}_{j_{m'} k_m}^{(\ell-1)}+{\bf q}_{\sf a}^{\sf T}\sum_{\substack{i=1\\i\neq k}}^K {\bf c}_{n_m i_m}^{(\ell-1)} +{\bf q}_{\sf r}^{\sf T} \sum_{\substack{m'=1\\m'\neq m}}^M\sum_{\substack{i=1\\i\neq k}}^K {\bf c}_{n_m i_{m'}}^{(\ell-1)} \Bigg), \notag
\end{equation}
where ${\bf p}_{\sf a}$ and ${\bf p}_{\sf r}$ are the weights used to aggregate information of intra-cell and inter-cell edges connected to UE$_{k_m}$, and ${\bf q}_{\sf a}$ and ${\bf q}_{\sf r}$ are the weights used to aggregate information of intra-cell and inter-cell edges connected to the $n$th antenna in the $m$th cell. We can again omit the aggregated information of edges not neighbored to edge $(n_m, k_m)$ as omitting term $(c)$ in \eqref{eq: upd propose GNN} to reduce the computational complexity of the Model-GNN.

\vspace{-2mm}\begin{remark}\emph{
		The proposed Model-GNN can be used to learn precoding policies that are optimized toward different objectives under different constraints. It is also applicable to learn hybrid precoding policies in mmWave or IRS systems, or to learn other matrix inverse-related policies such as channel estimation \cite{ChannelEst} or combiner in MIMO systems \cite{serbetli2004transceiver}. The differences of the Model-GNN for different policies lie in the inner product for computing $a_{jk}^{(\ell)}$, parameter sharing, and activation function of the output layer to satisfy constraints. Specifically, in a single-cell system, $a_{jk}^{(\ell)}={\bf x}_{j}^{\sf H}{\bf d}_k^{(\ell)}$, where ${\bf x}_{j}$ is the $j$th row of  matrix $\bf X$ whose pseudo-inverse is related to the concerned policy. For example, ${\bf X}$ is the channel matrix for precoding or combining optimization, and  $\bf X$ is the received pilot signal for channel estimation in MIMO systems.
	}	
\end{remark}

\vspace{-5mm}
\section{Simulation Results}\label{sec: simulations} 
In this section, we evaluate the performance of the proposed Model-GNN for learning precoding policies by comparing it with relevant baselines. We first consider a single-cell scenario, where SE-maximal and EE-maximal precoding policies are learned. Then, we provide the performance for learning the SE-maximal coordinated beamforming policy.

\vspace{-3mm}\subsection{Sample Generation and Hyper-parameters}
The  training and test samples are generated from uncorrelated Rayleigh fading channel, i.e., the elements in ${\bf H}$ follow complex Gaussian distribution ${\cal CN}(0, 1)$. The dimension of each sample of ${\bf H}$ depends on the considered settings of $N$ and $K$.
We generate 100,000 samples for every pair of $N$ and $K$ as the training set (the samples actually used for training are selected from the dataset, which may be with a much smaller number), and another 100 samples as the test set.

The hyper-parameters of the GNNs are shown in Table \ref{table: hyper params}. In the table, the number of neurons of each hidden layer (say the $\ell$th layer) is the dimension of the output vector of each edge (i.e., $J_{\ell}$), which is independent of $K$ and $N$ and hence can be used for different $K$ and $N$. The activation function of each layer is set as $\sigma({\bf X})={\bf X}/\|{\bf X}\|$ to avoid gradient diminishing. The DNNs are trained with unsupervised learning by Adam algorithm \cite{dlbook}.

\vspace{2mm}\begin{table}[!htb]
	\renewcommand{\arraystretch}{1.0}
	\centering
	\caption{Hyper-parameters for GNNs.}\label{table: hyper params}
	\small
\vspace{-3mm}
	\begin{tabular}{c|c|c|c|c}
		\hline\hline
		\multicolumn{2}{c|}{\bf Network} & \tabincell{c}{Num. of hidden layers} & \tabincell{c}{Num. of neurons in layers} & \tabincell{c}{Learning rate}  \\ \hline
		\multicolumn{2}{c|}{\bf Model-GNN (for maximizing SE)}							   &	3	& [32, 32, 8]& 0.01			\\ \hline
		\multicolumn{2}{c|}{\bf Model-GNN (for maximizing EE)}							   &	4	& [32, 32, 32, 8]	& 0.001*				\\ \hline	
		\multicolumn{2}{c|}{\bf Vanilla-GNN}							   &	4	& [64, 512, 512, 64]	& 0.01			\\ \hline
		\hline
	\end{tabular}
	\begin{flushleft}
		{\footnotesize 		
			*: We use a smaller learning rate for the Model-GNN to learn the EE-maximal precoding policy, such that the training is more stable and the trained GNN can better satisfy the constraints.} 
	\end{flushleft}\vspace{-1mm}
\end{table}

The performance of each DNN is obtained by averaging the test results of five independently trained DNNs to reduce the impact of randomness caused by the sample selection and initial weights. Each trained DNN is obtained with $N_{\sf tr}$ samples randomly selected from the training set and then tested on 100 samples. All the simulation results are obtained on a computer with one 14-core Intel i9-9940X CPU, one Nvidia RTX 2080Ti GPU, and 64 GB memory.

\vspace{-3mm}\subsection{Learning Precoding Policies in Single-cell Systems}

\subsubsection{Learning to maximize SE}\label{sec: simu result max SE}
We first validate Proposition \ref{prop: GNN element wise policy} that a Vanilla-GNN can only learn element-wise functions when $N$ and $K$ are large. To this end, we train the Vanilla-GNN to maximize SE with the samples generated in scenarios with different values of $N$ and $K$, and then compute the normalized correlation between the conjugated channel vector and the learned precoding vector of each user, say ${\bf h}_k^{\sf H}{\bf v}_k/(\|{\bf h}_k\|\cdot\|{\bf v}_k\|)$ for the $k$th user. If the correlation is $1$, then the Vanilla-GNN learns a MRT policy that is an element-wise function of $\bf H$.

\begin{figure}[!htb]
	\centering
	\begin{minipage}[t]{0.43\linewidth}	
		\subfigure[$N/K=2$]{
			\includegraphics[width=\textwidth]{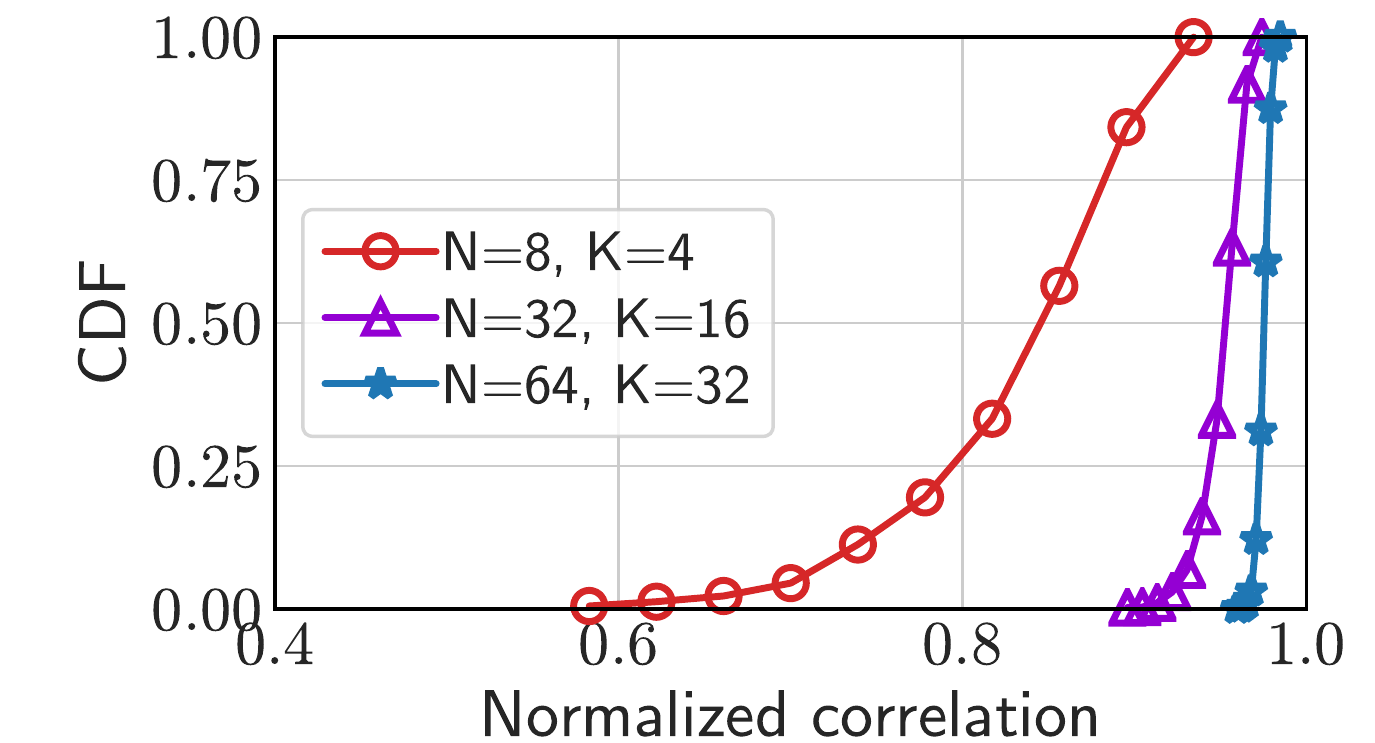}}
	\end{minipage} \hspace{5mm}
	\begin{minipage}[t]{0.43\linewidth}	
		\subfigure[$N/K=4$]{
			\includegraphics[width=\textwidth]{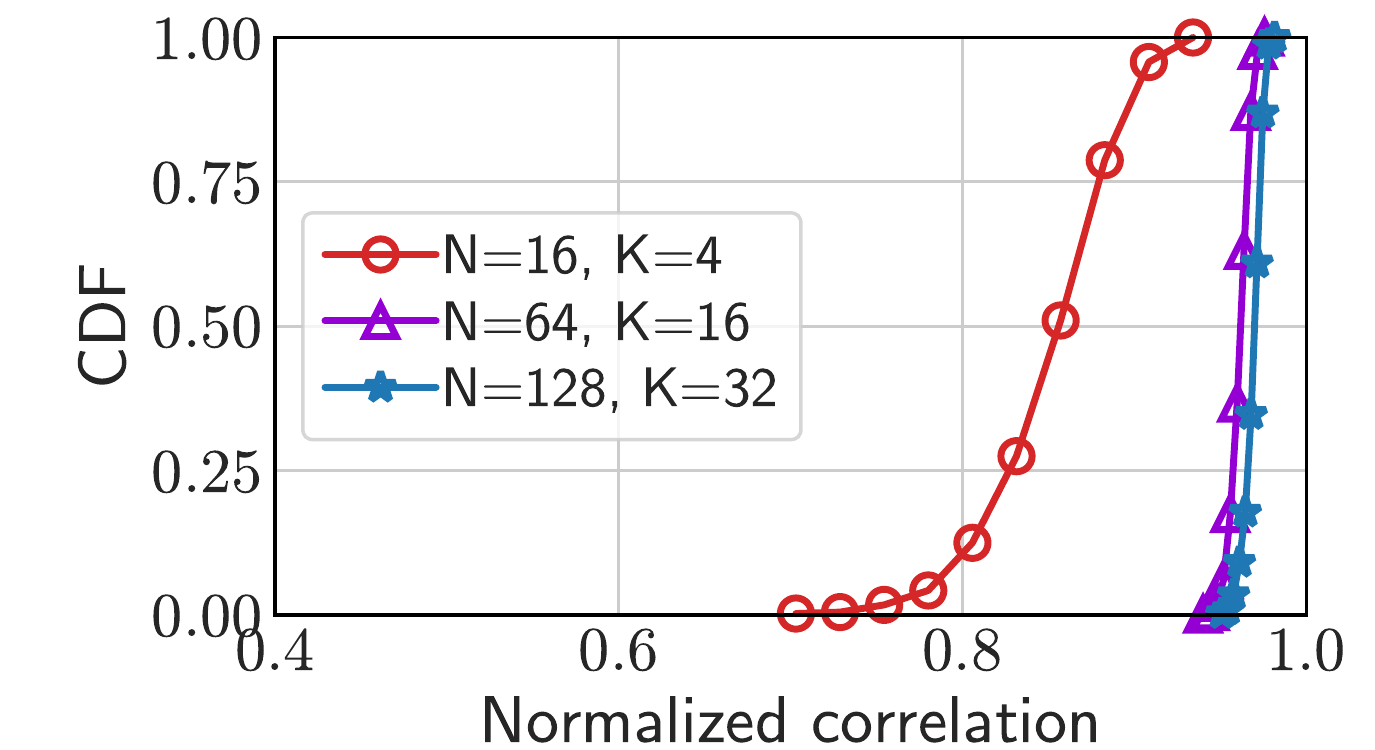}}
	\end{minipage}
	\vspace{-3mm}
	\caption{CDF of normalized correlation between channel vectors and learned precoding vectors of Vanilla-GNN, SNR=10 dB.}\label{fig-corr}
\end{figure}

In Fig. \ref{fig-corr}, we  provide the cumulative distribution functions (CDFs) of the normalized correlations on the test set. It is shown that when $N$ and $K$ are larger, the distribution becomes more concentrated on $1$. The normalized correlation also tends to be larger when $N/K$ is larger. This is because the channel vectors of the users tend to be mutually orthogonal when $N \gg K$,  such that the Vanilla-GNN tends to learn a MRT policy.

In what follows, we evaluate the performance of the Model-GNN and Vanilla-GNN for learning the precoding policy from problem \emph{P1}. As a comparison, we also provide the performance of ZFBF with equal power allocation (this is equivalent to using the well-trained TGNN) as well as two other model-driven DNNs in the literature. Since problem \emph{P1} can be solved via WMMSE algorithm \cite{WMMSE2011Shi}, the first model-driven DNN is the deep unfolding network proposed in \cite{DeepUnfold_WMMSE_TWC_2021}, where the WMMSE algorithm is unfolded into a multi-layer structure and only some parameters and operations are learned by neural networks. The number of layers of the unfolded network is set as five such that the time consumed for training is affordable, and other hyper-parameters are the same as in \cite{DeepUnfold_WMMSE_TWC_2021}. The second is the beamforming neural network (BNN) proposed in \cite{DL_Beamforming_model_TCOM2020}, where  the optimal solution structure of precoding in \eqref{eq: opt precod struc} is used to recover the precoding matrix with the learned power allocation. For a fair comparison, we use the Vanilla-GNN with three {hidden layers and $J_1=J_2=J_3=64$} to learn the power allocation instead of the CNN used in \cite{DL_Beamforming_model_TCOM2020}.
The performance is measured with the SE ratio, which is the SE achieved by the policy learned by each DNN divided by the SE achieved by the WMMSE algorithm.

\begin{figure}[!htb]
	\centering
	\begin{minipage}[t]{0.31\linewidth}	
		\subfigure[SNR=0 dB]{
			\includegraphics[width=\textwidth]{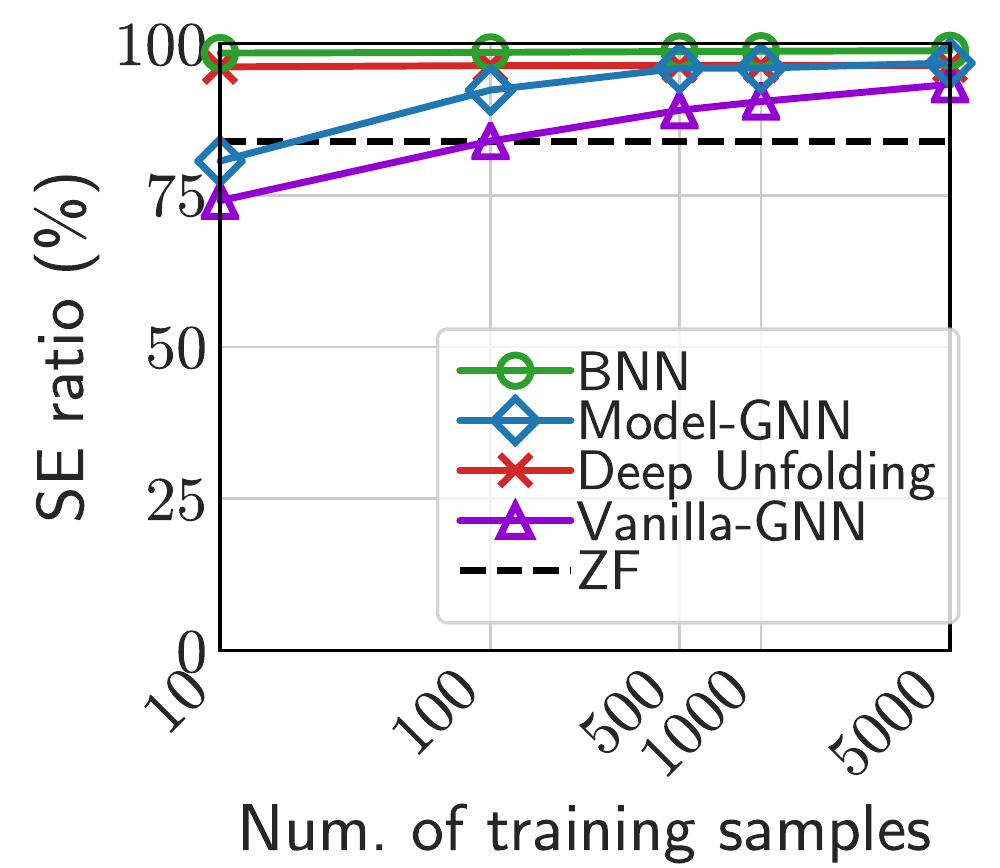}}
	\end{minipage}
	\begin{minipage}[t]{0.31\linewidth}	
		\subfigure[SNR=10 dB]{
			\includegraphics[width=\textwidth]{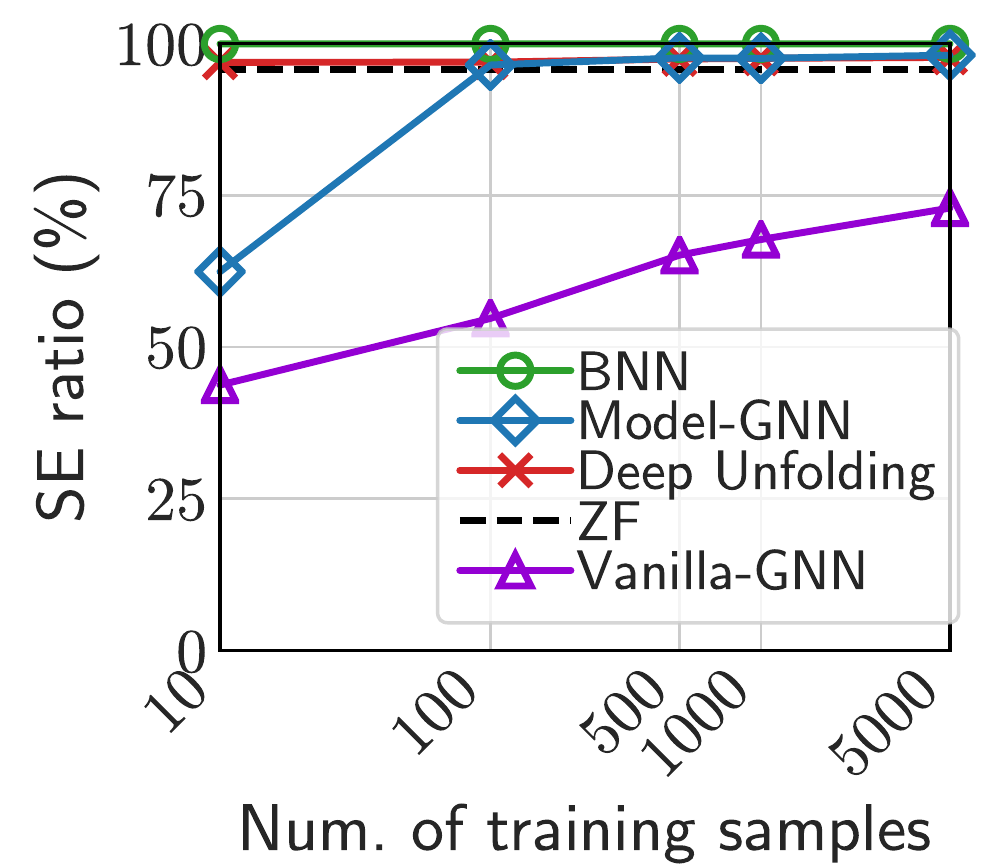}}
	\end{minipage}
	\begin{minipage}[t]{0.31\linewidth}	
		\subfigure[SNR=20 dB]{
			\includegraphics[width=\textwidth]{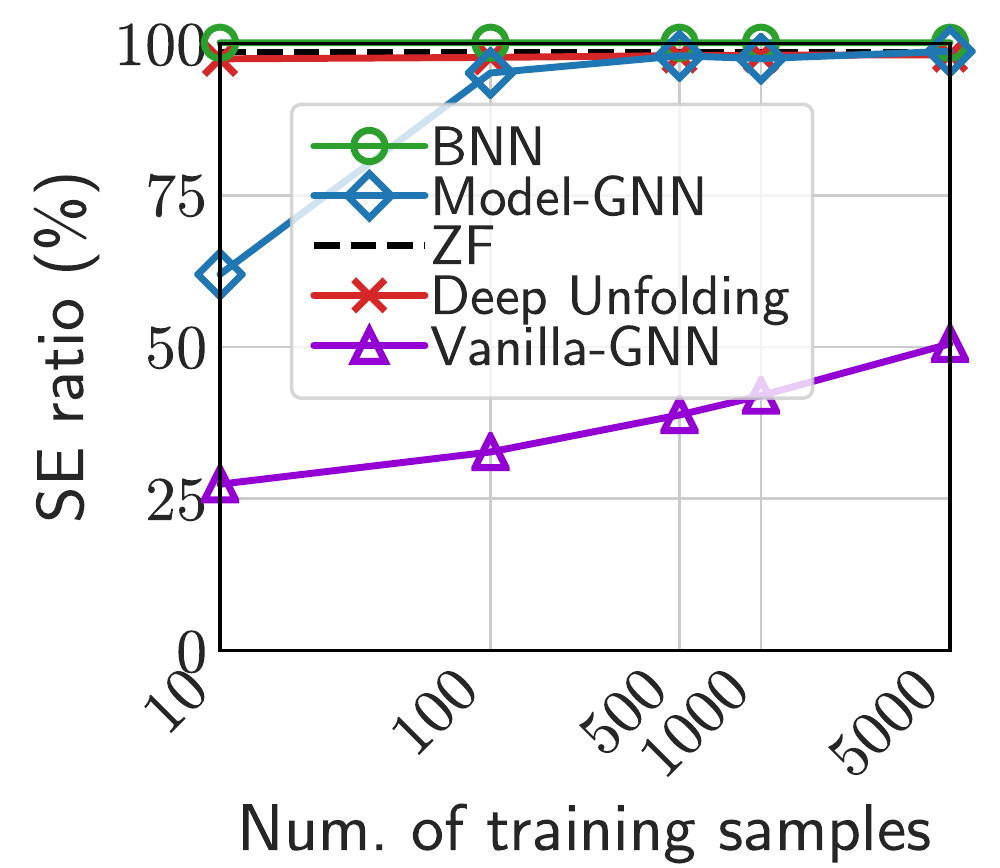}}
	\end{minipage}
	\vspace{-2mm}
	\caption{Learning performance of the GNNs and two existing model-driven DNNs, $\!N\!\!\!=\!\!8,\!K\!\!\!=\!\!4.\!$ The SE ratio of WMMSE is 100\%.}\label{fig-SE-Ntr-N8K4}\vspace{-1mm}
\end{figure}

In Fig. \ref{fig-SE-Ntr-N8K4}, we show the SE ratios under different SNRs. It can be seen that the Model-GNN can achieve higher than $95\%$ SE ratio at all SNRs with only 100 samples. The performance gap between the Model-GNN and Vanilla-GNN becomes smaller in a low SNR regime, because the Vanilla-GNN can learn a MRT policy that can achieve near-optimal performance in a noise-limited scenario.
The SE ratio of the Vanilla-GNN at high SNR can be improved by using more hidden layers and more neurons in each hidden layer and by using more training samples, but the results are not shown since the number of required training samples grow with SNR rapidly.
The  SE ratios achieved by the deep unfolding network and the BNN are close to 100\% with only 10 samples.

In Table \ref{table: compu cplxty SE}, we compare the SE ratios and the inference time of the DNNs and the WMMSE algorithm under different values of $N, K$ and SNR. The SE ratios and the inference time are respectively averaged over 100 test samples. All the DNNs are trained with 1,000 samples. The inference time is measured by the time of running the trained Model-GNN, Vanilla-GNN and BNN on GPU. Since the code of the deep unfolding network provided in \cite{DeepUnfold_WMMSE_TWC_2021} and the WMMSE algorithm can only run on CPU, we also measure the inference time of the Model-GNN on CPU, such that its computational complexity is comparable with these two methods. We can see that the Model-GNN can achieve higher than 97\% SE ratio for all scenarios. By contrast, the SE ratios of the Vanilla-GNN are much lower, especially when $N$ and $K$ are large and SNR is high.
The inference time of the Model-GNN on GPU only grows slightly with the problem scale, and is lower than the Vanilla-GNN, which seems contradicted with the FLOPs calculated at section \ref{sec: cpt cplxty}.
This is because the number of neurons of the Vanilla-GNN in each layer is much higher than the Model-GNN when learning the precoding policies, as given in Table \ref{table: hyper params}. The inference time of the Model-GNN on CPU grows with the problem scale, but is still much lower than the time consumed by the deep unfolding network (which involves computational burdened operations such as matrix multiplications over iterations) and the WMMSE algorithm.
The SE achieved by the BNN is close to or even higher than the WMMSE algorithm, with slightly shorter running time than the Model-GNN.

\vspace{2mm}
\begin{table}[!htb]
	\renewcommand{\arraystretch}{1.05}
	\centering
	\footnotesize
	\caption{SE ratios and Inference Time (in seconds), 1000 samples are used for training each DNN}\vspace{-3mm} \label{table: compu cplxty SE}
		\begin{tabular}{c|c|ccc|cc|cc|cc|cc}
		\hline\hline
		\multirow{2}{*}{\tabincell{c}{\bf SNR\\\bf (dB)}} & \multicolumn{1}{c|}{\multirow{2}{*}{$(N, K)$}} & \multicolumn{3}{c|}{\bf Model-GNN}                                               & \multicolumn{2}{c|}{\bf Vanilla-GNN}     & \multicolumn{2}{c|}{\bf Deep unfolding*} & \multicolumn{2}{c|}{\bf BNN}              & \multicolumn{2}{c}{\bf WMMSE}             \\ \cline{3-13}
		& \multicolumn{1}{c|}{}                        & \multicolumn{1}{c|}{\tabincell{c}{\bf SE\\\bf ratio}} & \multicolumn{1}{c|}{\tabincell{c}{\bf Time\\\bf (GPU)}} & \tabincell{c}{\bf Time\\\bf (CPU)} & \multicolumn{1}{c|}{\tabincell{c}{\bf SE\\\bf ratio}} & \tabincell{c}{\bf Time\\\bf (GPU)} & \multicolumn{1}{c|}{\tabincell{c}{\bf SE\\\bf ratio}} & \tabincell{c}{\bf Time\\\bf (CPU)}  & \multicolumn{1}{c|}{\tabincell{c}{\bf SE\\\bf ratio}} & \tabincell{c}{\bf Time\\\bf (GPU)} & \multicolumn{1}{c|}{\tabincell{c}{\bf SE\\\bf ratio}} & \tabincell{c}{\bf Time\\\bf (CPU)}   \\ \hline
		\multirow{5}{*}{0}          & (8, 4)   & \multicolumn{1}{c|}{97.76}    & \multicolumn{1}{c|}{0.06}       &   0.12         & \multicolumn{1}{c|}{90.42}    & 0.19 & \multicolumn{1}{c|}{96.37}    & 1.07  & \multicolumn{1}{c|}{98.74}    & 0.03 & \multicolumn{1}{c|}{100}      & 0.9    \\ \cline{2-13}
		& (32, 16)                                     & \multicolumn{1}{c|}{99.01}    & \multicolumn{1}{c|}{0.07}       &  0.13          & \multicolumn{1}{c|}{81.63}      & 0.19  & \multicolumn{1}{c|}{95.08}      & 7.89   & \multicolumn{1}{c|}{99.07}    & 0.03 & \multicolumn{1}{c|}{100}      & 6.85   \\ \cline{2-13}
		& (64, 16)                                     & \multicolumn{1}{c|}{99.74}    & \multicolumn{1}{c|}{0.08}        &  0.20          & \multicolumn{1}{c|}{69.71}      & 0.17  & \multicolumn{1}{c|}{92.67}      & 8.73   & \multicolumn{1}{c|}{95.55}    & 0.04 & \multicolumn{1}{c|}{100}      & 39.78  \\ \cline{2-13}
		& (64, 32)                                     & \multicolumn{1}{c|}{99.42}    & \multicolumn{1}{c|}{0.09}       &  0.21          & \multicolumn{1}{c|}{68.37}      & 0.18  & \multicolumn{1}{c|}{96.04}      &  {28.89}  & \multicolumn{1}{c|}{99.57}    & 0.05 & \multicolumn{1}{c|}{100}      & 51.3   \\ \hline
		\multirow{5}{*}{10}         & (8, 4)  & \multicolumn{1}{c|}{97.74}    & \multicolumn{1}{c|}{0.06}       &  0.12          & \multicolumn{1}{c|}{67.7}     & 0.17 & \multicolumn{1}{c|}{97.69}    & 1.08 & \multicolumn{1}{c|}{99.92}    & 0.04 & \multicolumn{1}{c|}{100}      & 1.37   \\ \cline{2-13}
		& (32, 16)                                     & \multicolumn{1}{c|}{99.03}    & \multicolumn{1}{c|}{0.07}       &  0.13          & \multicolumn{1}{c|}{42.74}      & 0.17  & \multicolumn{1}{c|}{90.64}      & 7.23   & \multicolumn{1}{c|}{99.97}    & 0.04 & \multicolumn{1}{c|}{100}      & 12.22  \\ \cline{2-13}
		& (64, 16)                                     & \multicolumn{1}{c|}{99.65}    & \multicolumn{1}{c|}{0.08}       &  0.20          & \multicolumn{1}{c|}{42.05}      & 0.17  & \multicolumn{1}{c|}{87.97}      & 8.74   & \multicolumn{1}{c|}{99.99}    & 0.03 & \multicolumn{1}{c|}{100}      & 122.25 \\ \cline{2-13}
		& (64, 32)                                     & \multicolumn{1}{c|}{99.41}    & \multicolumn{1}{c|}{0.09}       & 0.21           & \multicolumn{1}{c|}{37.45}      & 0.18  & \multicolumn{1}{c|}{80.38}      & {28.51}   & \multicolumn{1}{c|}{100.4}    & 0.05  & \multicolumn{1}{c|}{100}      & 93.05  \\ \hline
		\multirow{5}{*}{20}         & (8, 4)  & \multicolumn{1}{c|}{97.49}    & \multicolumn{1}{c|}{0.06}       &  0.13          & \multicolumn{1}{c|}{38.81}    & 0.16 & \multicolumn{1}{c|}{98.11}    & 1.08 & \multicolumn{1}{c|}{100.1}    & 0.04  & \multicolumn{1}{c|}{100}      & 5.6    \\ \cline{2-13}
		& (32, 16)                                     & \multicolumn{1}{c|}{98.35}    & \multicolumn{1}{c|}{0.08}       &  0.13          & \multicolumn{1}{c|}{24.23}     & 0.16 & \multicolumn{1}{c|}{75.87}      & 7.96   & \multicolumn{1}{c|}{101.19}   & 0.04 & \multicolumn{1}{c|}{100}      & 43.96  \\ \cline{2-13}
		& (64, 16)                                     & \multicolumn{1}{c|}{99.75}    & \multicolumn{1}{c|}{0.08}       &  0.20          & \multicolumn{1}{c|}{28.16}      & 0.17  & \multicolumn{1}{c|}{83.66}      & 8.81   & \multicolumn{1}{c|}{100.08}   & 0.04 & \multicolumn{1}{c|}{100}      & 289.41 \\ \cline{2-13}
		& (64, 32)                                     & \multicolumn{1}{c|}{98.94}    & \multicolumn{1}{c|}{0.09}       &  0.21          & \multicolumn{1}{c|}{25.32}      & 0.18  & \multicolumn{1}{c|}{77.47}      & {28.89}   & \multicolumn{1}{c|}{101.67}   & 0.05 & \multicolumn{1}{c|}{100}      & 276.87 \\ \hline\hline
	\end{tabular}
	\begin{flushleft}
		{\footnotesize 		
			*: For the deep unfolding network, we only train it for 10 epochs such that the training time is affordable. Hence, the SE ratio is low for large $N, K$ and high SNR. }
	\end{flushleft}
\vspace{-3mm}
\end{table}

In Table \ref{table: train cplxty}, we compare the sample and time complexities for training the DNNs, which are defined as the minimal number of training samples and minimal training time for achieving an expected performance (set as 95\% SE ratio). For the same reason mentioned above, the training time of DNNs with GPU or CPU or both of them is given. As expected, the BNN has the lowest training complexity, which only needs to learn the power allocated to the $K$ users. The Model-GNN also needs low training complexity. Although the sample complexity of the deep unfolding network is lower than the Model-GNN, the time complexity of the deep unfolding network is much higher, due to the computation-burdened operations. Both the sample and time complexities of the Vanilla-GNN are much higher than the Model-GNN.

\vspace{2mm}\begin{table}[!htb]
	\renewcommand{\arraystretch}{1.0}
	\centering
	\small
	\vspace{1mm}
	\caption{Sample Complexity and Training Time, SNR=10 dB}\vspace{-1mm} \label{table: train cplxty}
	\vspace{-2mm}
\begin{tabular}{c|c|c|c|c|c}
	\hline \hline
	&              & \bf Model-GNN & \bf Vanilla-GNN & \bf BNN & \bf Deep Unfolding \\ \hline
	\multirow{3}{*}{$N=8, K=4$}   & Sample       & 100       & 40,000   &  10   & 10             \\ \cline{2-6}
	& CPU time (s) & 1 min 2 s     & ---     & ---    & 4 min 43 s         \\ \cline{2-6}
	& GPU time (s) & 32 s     & 43 min 16 s  & 11 s  & ---           \\ \hline
	\multirow{3}{*}{$N=32, K=16$} & Sample       & 100       &   $>$100,000      & 10   & 10            \\ \cline{2-6}
	& CPU time (s) & 4 min 35 s          & ---      & ---   & 70 min 21 s           \\ \cline{2-6}
	& GPU time (s) & 39 s     &  $>$6 hours 22 min 48 s   & 21 s       & ---            \\ \hline \hline
\end{tabular}
\vspace{-1mm}
\end{table}

\begin{figure}[!htb]
	\centering
	\includegraphics[width=0.75\linewidth]{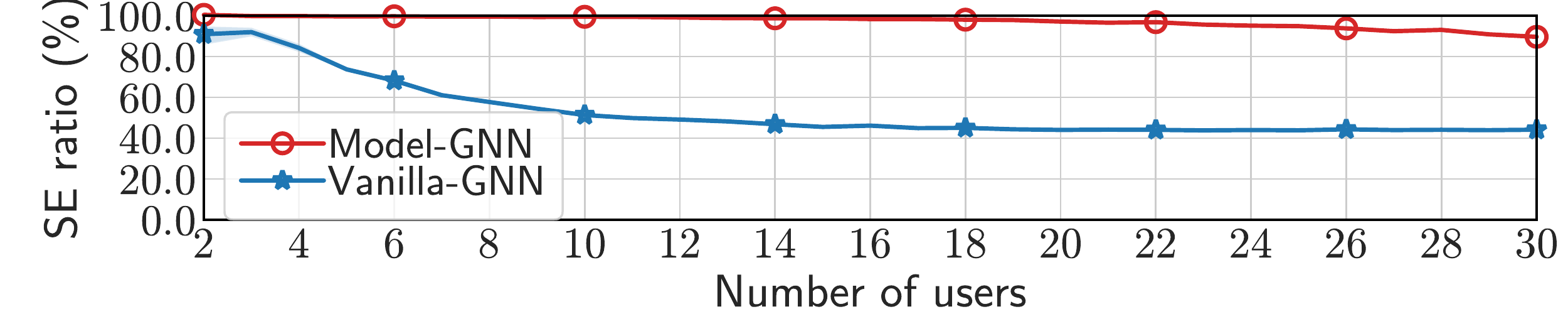}\vspace{-3mm}
	\caption{Generalization ability of the GNNs to $K$, $N=32$, SNR=10 dB.}
	\label{fig-scale-SE}
\end{figure}

In Fig. \ref{fig-scale-SE}, we show the SE ratios achieved by the GNNs tested in the scenarios with different values of $K$ from the training samples. The GNNs are trained with 1,000 samples, which are generated in scenarios where $K$ follows exponential distribution with mean value of $4$. In this case, $96$\% of samples are generated for the case with $K<10$, hence the GNNs can be trained in small problem scales with low complexity. The test samples are generated in scenarios where $K\thicksim{\mathbb U}(2,30)$, where ${\mathbb U}(\cdot,\cdot)$ stands for uniform distribution.
 It can be seen that the SE ratios of the Model-GNN degrade very slowly, which indicates its good generalization ability to $K$.

\vspace{2mm}

In previous simulations, the noise power $\sigma_0^2$ and the maximal transmit power $P_{\max}$ are fixed, which are not inputted into the Model-GNN. To allow the Model-GNN to perform well under different SNRs, we input  ${\bf H}'=[{h_{nk}}']^{N\times K}$ into the Model-GNN, where ${{h_{nk}}'}=h_{nk}\sqrt{P_{\max}/\sigma_0^2}$. Our simulation results show that a Model-GNN trained with the samples generated with the SNRs randomly selected from $[0, 20]$ dB performs well in the test samples generated with the SNRs in the same range. Due to the limited space, the results are not provided.

\subsubsection{Learning to maximize EE}
To demonstrate that the proposed Model-GNN can be used to learn other precoding policies, we train the DNNs to maximize EE (defined as $\big({\sum_{k=1}^K R_k({\bf H, V})}\big)/\big({\rho {\sf Tr}({\bf V}^{\sf H}{\bf V}) + NP_c + P_0}\big)$) with quality of service (QoS) constraints  $R_k({\bf H, V}) \geq R_{\min}, k=1,\cdots,K$, where $P_c$  is the circuit power consumption for each antenna, $P_0$ is the constant power consumption for each BS, $1/\rho$ is the power amplifier efficiency, and $R_{\min}$ is the minimal data rate requirement of each user. In the simulation, $P_c=17.6$ W, $P_0=43.3$ W, $1/\rho=0.311$, which come from \cite{auer2011much} for a macro BS.

The EE-maximal problem can be solved with the iterative algorithm proposed in \cite{LY_TVT_2015}, which is referred to as \emph{EE-numerical} in the sequel. We set the number of iterations as 100 such that the computation time is affordable.
The algorithm may only find sub-optimal solutions after iterations due to the non-convexity of the problem.

We train the DNNs with unsupervised learning. Since the QoS constraints cannot be satisfied by passing through a normalization layer as for problem \emph{P1}, we resort to the Lagrangian multiplier method \cite{sun2019pimrc}, which strives to satisfy the QoS requirements during training but cannot guarantee the constraints to be  satisfied. Hence, we measure the learning performance in terms of EE and QoS satisfaction by \emph{EE ratio} and \emph{Constraint Satisfaction Ratio}. The EE ratio is defined as the ratio of EE achieved by the learned policy to the EE achieved by \emph{EE-numerical}. The constraint satisfaction ratio is  the percentage of the users whose QoS requirements are satisfied (averaged over all the test samples), i.e.,
\begin{equation}
{\sf Constraint~Satisfaction~Ratio} = \frac{\sum_{i=1}^{N_{\sf te}}\sum_{k=1}^K {\bf 1}\big(R_k({\bf H}_{[i]}, \hat{\bf V}_{[i]})\geq R_{\min}\big)}{N_{\sf te} K}, \notag
\end{equation}
where $N_{\sf te}$ is the number of test samples, the subscript ${[i]}$ indicates the $i$th sample, ${\bf 1}(x)=1$ if $x$ is true and ${\bf 1}(x)=0$ otherwise.

In Table \ref{table-ee}, we compare the EE ratios, constraint satisfaction ratios and the inference time of the DNNs and the numerical algorithms over 100 test samples under different values of $N, K$ and $R_{\min}$ (both ratios of the \emph{EE-numerical} algorithm are 100\%, and the running time of WMMSE algorithm has been provided in Table \ref{table: compu cplxty SE}, and hence is no longer listed). All the DNNs are trained with 10,000 samples, which is more than the samples used to learn the SE-maximal precoding policy such that the QoS constraint can be better satisfied. The inference time of all the DNNs is measured on GPU. Since the \emph{EE-numerical}  algorithm can only run on CPU, we also measure the inference time of the Model-GNN on CPU. We can see that the Model-GNN can achieve over 90\% EE ratio and constraint satisfaction ratio in all scenarios, which are much higher than the Vanilla-GNN.  The inference time of the Model-GNN is comparable to the Vanilla-GNN and BNN, and is much lower than the \emph{EE-numerical}  algorithm. The BNN can also achieve high EE ratios for all scenarios, but the constraint satisfaction ratio becomes lower when $R_{\min}$ is larger. The EE ratio achieved by the WMMSE algorithm is very low, and the constraint satisfaction ratio is zero when $R_{\min}$ is high, because this algorithm is designed to maximize SE with power constraints. Hence, the deep unfolding network in \cite{DeepUnfold_WMMSE_TWC_2021} also cannot learn the EE-maximal policies (not shown for conciseness).

\begin{table}[!htb]
	\renewcommand{\arraystretch}{1.05}
	\centering
	\footnotesize
	\caption{EE Ratio (EER),  constraint satisfaction ratio (CSR) and Running Time (in seconds), SNR=10 dB}\label{table-ee}
\vspace{-2mm}	\setlength\tabcolsep{3pt}
\begin{tabular}{c|c|cccc|ccc|ccc|cc|c}
	\hline\hline
	\multirow{3}{*}{\tabincell{c}{$R_{\min}$\\(bps)}} & \multirow{2}{*}{$(N, K)$} & \multicolumn{4}{c|}{\textbf{Model-GNN}}                                                                                                                                                                                                                                                                                                      & \multicolumn{3}{c|}{\textbf{Vanilla-GNN}}                                                                                                                                                                                                          & \multicolumn{3}{c|}{\textbf{BNN}}                                                                                                                                                                                                                  & \multicolumn{2}{c|}{\textbf{WMMSE}}                                                                                                                      & \multicolumn{1}{c}{\tabincell{c}{\bf EE-Numerical}}                            \\ \cline{3-15}
	&                         & \multicolumn{1}{c|}{\textbf{\begin{tabular}[c]{@{}c@{}}EER\\      (\%)\end{tabular}}} & \multicolumn{1}{c|}{\textbf{\begin{tabular}[c]{@{}c@{}}CSR\\      (\%)\end{tabular}}} & \multicolumn{1}{c|}{\textbf{\begin{tabular}[c]{@{}c@{}}Time\\      (GPU)\end{tabular}}} & \textbf{\begin{tabular}[c]{@{}c@{}}Time\\      (CPU)\end{tabular}} & \multicolumn{1}{c|}{\textbf{\begin{tabular}[c]{@{}c@{}}EER\\      (\%)\end{tabular}}} & \multicolumn{1}{c|}{\textbf{\begin{tabular}[c]{@{}c@{}}CSR\\      (\%)\end{tabular}}} & \textbf{\begin{tabular}[c]{@{}c@{}}Time\\      (GPU)\end{tabular}} & \multicolumn{1}{c|}{\textbf{\begin{tabular}[c]{@{}c@{}}EER\\      (\%)\end{tabular}}} & \multicolumn{1}{c|}{\textbf{\begin{tabular}[c]{@{}c@{}}CSR\\      (\%)\end{tabular}}} & \textbf{\begin{tabular}[c]{@{}c@{}}Time\\      (GPU)\end{tabular}} & \multicolumn{1}{c|}{\textbf{\begin{tabular}[c]{@{}c@{}}EER\\      (\%)\end{tabular}}} & \textbf{\begin{tabular}[c]{@{}c@{}}CSR\\      (\%)\end{tabular}} & \textbf{\begin{tabular}[c]{@{}c@{}}Time\\      (CPU)\end{tabular}} \\ \hline
	\multirow{3}{*}{3}    & (8, 4)                  & \multicolumn{1}{c|}{97.29}                                                            & \multicolumn{1}{c|}{100}                                                              & \multicolumn{1}{c|}{0.07}                                                               & 0.14                                                               & \multicolumn{1}{c|}{47.84}                                                            & \multicolumn{1}{c|}{62.25}                                                            & 0.05                                                               & \multicolumn{1}{c|}{100.23}                                                           & \multicolumn{1}{c|}{100}                                                              & 0.05                                                               & \multicolumn{1}{c|}{63.18}                                                            & 88.75                                                            & 18.74                                                              \\ \cline{2-15}
	& (32, 16)                & \multicolumn{1}{c|}{95.93}                                                            & \multicolumn{1}{c|}{100}                                                              & \multicolumn{1}{c|}{0.07}                                                               & 0.15                                                               & \multicolumn{1}{c|}{28.12}                                                            & \multicolumn{1}{c|}{1.88}                                                             & 0.06                                                               & \multicolumn{1}{c|}{96.87}                                                            & \multicolumn{1}{c|}{100}                                                              & 0.05                                                               & \multicolumn{1}{c|}{52.43}                                                            & 93.31                                                            & 144.38                                                             \\ \cline{2-15}
	& (32, 24)                & \multicolumn{1}{c|}{101.6}                                                            & \multicolumn{1}{c|}{99.99}                                                            & \multicolumn{1}{c|}{0.09}                                                               & 0.15                                                               & \multicolumn{1}{c|}{20.98}                                                            & \multicolumn{1}{c|}{0}                                                                & 0.07                                                               & \multicolumn{1}{c|}{100.69}                                                           & \multicolumn{1}{c|}{100}                                                              & 0.08                                                               & \multicolumn{1}{c|}{51.18}                                                            & 43.04                                                            & 1313.8                                                             \\ \hline
	\multirow{3}{*}{5}    & (8, 4)                  & \multicolumn{1}{c|}{95.5}                                                            & \multicolumn{1}{c|}{98.5}                                                            & \multicolumn{1}{c|}{0.07}                                                               & 0.14                                                               & \multicolumn{1}{c|}{47.9}                                                             & \multicolumn{1}{c|}{2}                                                                & 0.05                                                               & \multicolumn{1}{c|}{100.1}                                                            & \multicolumn{1}{c|}{100}                                                              & 0.05                                                               & \multicolumn{1}{c|}{62.41}                                                            & 1.5                                                              & 18.55                                                              \\ \cline{2-15}
	& (32, 16)                & \multicolumn{1}{c|}{95.97}                                                            & \multicolumn{1}{c|}{100}                                                              & \multicolumn{1}{c|}{0.07}                                                               & 0.15                                                               & \multicolumn{1}{c|}{26.04}                                                            & \multicolumn{1}{c|}{0}                                                                & 0.06                                                               & \multicolumn{1}{c|}{96.63}                                                            & \multicolumn{1}{c|}{100}                                                              & 0.06                                                               & \multicolumn{1}{c|}{52.41}                                                            & 0                                                                & 144.14                                                             \\ \cline{2-15}
	& (32, 24)                & \multicolumn{1}{c|}{96.32}                                                            & \multicolumn{1}{c|}{99.87}                                                            & \multicolumn{1}{c|}{0.07}                                                               & 0.15                                                               & \multicolumn{1}{c|}{21.5}                                                             & \multicolumn{1}{c|}{0}                                                                & 0.08                                                               & \multicolumn{1}{c|}{95.6}                                                             & \multicolumn{1}{c|}{100}                                                              & 0.08                                                               & \multicolumn{1}{c|}{48.46}                                                            & 0                                                                & 1347.8                                                             \\ \hline
	\multirow{3}{*}{7}    & (8, 4)                  & \multicolumn{1}{c|}{95.95}                                                            & \multicolumn{1}{c|}{98.83}                                                            & \multicolumn{1}{c|}{0.07}                                                               & 0.14                                                               & \multicolumn{1}{c|}{50.8}                                                            & \multicolumn{1}{c|}{0.25}                                                             & 0.05                                                               & \multicolumn{1}{c|}{98.7}                                                             & \multicolumn{1}{c|}{94.25}                                                            & 0.05                                                               & \multicolumn{1}{c|}{62.46}                                                            & 0                                                                & 11.46                                                              \\ \cline{2-15}
	& (32, 16)                & \multicolumn{1}{c|}{93.84}                                                            & \multicolumn{1}{c|}{100}                                                              & \multicolumn{1}{c|}{0.07}                                                               & 0.15                                                               & \multicolumn{1}{c|}{26.18}                                                            & \multicolumn{1}{c|}{0}                                                                & 0.06                                                               & \multicolumn{1}{c|}{94.25}                                                            & \multicolumn{1}{c|}{100}                                                              & 0.05                                                               & \multicolumn{1}{c|}{51.09}                                                            & 0                                                                & 144.48                                                             \\ \cline{2-15}
	& (32, 24)                & \multicolumn{1}{c|}{90.15}                                                            & \multicolumn{1}{c|}{97.44}                                                             & \multicolumn{1}{c|}{0.08}                                                               & 0.15                                                               & \multicolumn{1}{c|}{22.8}                                                            & \multicolumn{1}{c|}{0}                                                              & 0.08                                                               & \multicolumn{1}{c|}{89.27}                                                            & \multicolumn{1}{c|}{99.6}                                                             & 0.08                                                               & \multicolumn{1}{c|}{45.54}                                                            & 0                                                                & 1541.1                                                             \\ \hline
	\multirow{3}{*}{9}    & (8, 4)                  & \multicolumn{1}{c|}{96.57}                                                            & \multicolumn{1}{c|}{91.65}                                                            & \multicolumn{1}{c|}{0.07}                                                               & 0.14                                                               & \multicolumn{1}{c|}{59.55}                                                            & \multicolumn{1}{c|}{0}                                                                & 0.06                                                               & \multicolumn{1}{c|}{95.43}                                                            & \multicolumn{1}{c|}{77}                                                               & 0.05                                                               & \multicolumn{1}{c|}{74.59}                                                            & 0                                                                & 2.97                                                               \\ \cline{2-15}
	& (32, 16)                & \multicolumn{1}{c|}{91.37}                                                            & \multicolumn{1}{c|}{99.27}                                                            & \multicolumn{1}{c|}{0.07}                                                               & 0.15                                                               & \multicolumn{1}{c|}{25.78}                                                            & \multicolumn{1}{c|}{0}                                                             & 0.06                                                               & \multicolumn{1}{c|}{93.38}                                                            & \multicolumn{1}{c|}{99.69}                                                            & 0.05                                                               & \multicolumn{1}{c|}{50.81}                                                            & 0                                                                & 153.6                                                              \\ \cline{2-15}
	& (32, 24)                & \multicolumn{1}{c|}{90.09}                                                            & \multicolumn{1}{c|}{90.25}                                                              & \multicolumn{1}{c|}{0.1}                                                                & 0.15                                                               & \multicolumn{1}{c|}{28.23}                                                            & \multicolumn{1}{c|}{0}                                                                & 0.06                                                               & \multicolumn{1}{c|}{90.56}                                                            & \multicolumn{1}{c|}{85.56}                                                            & 0.08                                                               & \multicolumn{1}{c|}{56.75}                                                            & 0                                                                & 700.9                                                              \\ \hline\hline
\end{tabular}
\vspace{-1mm}
\end{table}

\begin{figure}[!htb]
	\centering
	\includegraphics[width=0.9\linewidth]{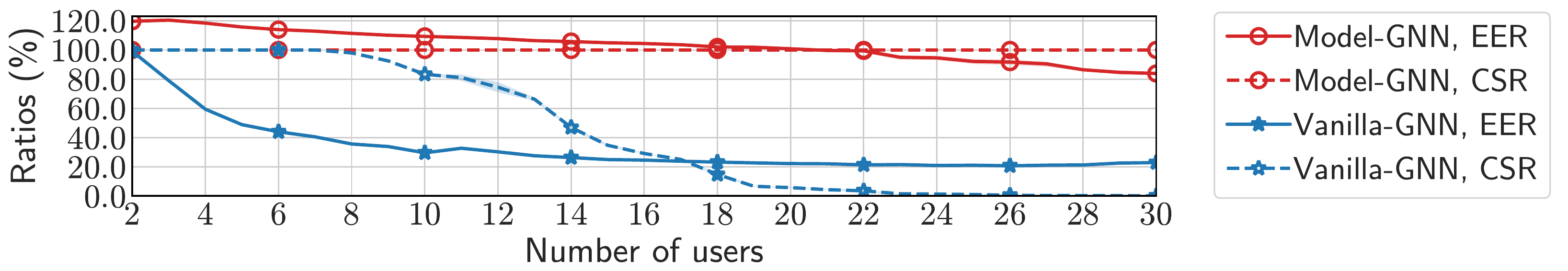}\vspace{-3mm}
	\caption{Generalization ability of the GNNs to $K$, $N=32$, $R_{\min}=2$ bits/Hz, SNR=10 dB.}
	\label{fig-scale-EE}
\end{figure}

In Fig. \ref{fig-scale-EE}, we show the EE ratio and constraint satisfaction ratio achieved by the GNNs tested in scenarios with different values of $K$ from the training samples. The GNNs are trained with 10,000 samples,
which are generated in scenarios where $K$ follows exponential distribution with mean value of 4,  and are tested with 1,000 samples generated in scenarios where $K\thicksim {\mathbb U}(2,30)$. Since the power consumption for maximizing EE varies with $K$, we train a FNN with $K$ as input and a factor $\eta_K$ as output, and then multiply $\eta_K$  with the learned precoding matrix of Model-GNN. The FNN is jointly trained with the Model-GNN. It can be seen that the two ratios achieved by the Model-GNN degrade slowly (or do not degrade) with $K$, which demonstrates its superior generalization ability. The EE ratios achieved by the GNNs exceed 100\% when $K$ is small, because the numerical algorithm only finds the suboptimal solutions.

\vspace{2mm}

\vspace{-4mm}\subsection{Learning Coordinated Beamforming Policy}
To evaluate the performance of the Model-GNN for learning the precoding policy to coordinate inter-cell interference, we consider problem \emph{P2} in subsection \ref{subsec: multicell}. This problem can also be solved numerically with the WMMSE algorithm.

\vspace{2mm}
\begin{table}[!htb]
	\renewcommand{\arraystretch}{1.0}
	\centering
	\caption{SE Ratios of the Learned Coordinated Beamforming, SNR=10 dB}\label{table: multi-cell}
	\vspace{-3mm} \small
	\begin{tabular}{c|c|c|c|c}
		\hline\hline
		\multicolumn{2}{c|}{\bf Network} & $M=2, N=16, K=8$ & $M=3, N=16, K=8$ & $M=3, N=8, K=4$  \\ \hline
		\multicolumn{2}{c|}{\bf Model-GNN}		& 98.88\%	& 97.33\%  &  	96.39\%		\\ \hline
		\multicolumn{2}{c|}{\bf Vanilla-GNN}	& 40.65\%	& 36.67\%  &  	35.7\%		\\ \hline
		\hline
	\end{tabular}
\end{table}

In Table. \ref{table: multi-cell}, we show the SE ratios achieved by the DNNs trained with 1,000 samples in different scenarios.
Since the structure in \eqref{eq: opt precod struc} is no longer optimal for coordinated beamforming, we only compare the performance of the Model-GNN with the Vanilla-GNN.
It can be seen that the Model-GNN can achieve over 95\% SE ratio, which is much higher than the SE ratios achieved by the Vanilla-GNN.
The performance of the Vanilla-GNN can be improved by training larger networks with more samples, but is still much worse than the Model-GNN even when it is trained with 100,000 samples.
The Model-GNN is also generalizable to different numbers of cells and users. The results are not provided due to the lack of space.

\vspace{-2mm}
\subsection{Summary of Major Results} \vspace{-1mm}
 Simulation results validate that the Model-GNN can be used to learn both SE-maximal and EE-maximal precoding policies under different constraints in single- and multi-cell systems.

1) In the single-cell system, we obtain the following observations.
\begin{itemize}
	\item The policy learned by the Model-GNN achieves much higher SE, and much higher EE with better satisfied QoS than the policy learned by the Vanilla-GNN with the same number of training samples. The Model-GNN needs much lower training complexity to achieve the same performance as the Vanilla-GNN.
	\item The policy learned by the Model-GNN achieves a comparable SE ratio to the policies learned by the deep unfolding network in \cite{DeepUnfold_WMMSE_TWC_2021} and BNN in \cite{DL_Beamforming_model_TCOM2020}, but with much lower computational complexities than the deep unfolding network in both training and inference phases  when learning the SE-maximal precoding. The deep unfolding network in \cite{DeepUnfold_WMMSE_TWC_2021} is not applicable for learning EE-maximal precoding policies. The Model-GNN can achieve comparable EE ratios to the BNN but can better satisfy the QoS requirements than BNN when learning the EE-maximal precoding.
	\item The trained Model-GNNs for learning both the SE- and EE-maximal policies can be generalized well to the number of users.
\end{itemize}

2) In the multi-cell system, the Model-GNN also performs well, whereas the BNN is not applicable to learn the coordinated beamforming policy.

\vspace{-2mm}\section{Conclusions}\label{sec: conclusions}
\vspace{-1mm}
In this paper, we proposed a Model-GNN to learn precoding policies in multi-antenna systems. Motivated by the fact that matrix inverse is a common operation in precoding to mitigate multi-user interference and is hard to learn, we resorted to the Taylor's expansion formula of matrix pseudo-inverse for designing a GNN that is neither algorithm-specific nor problem-specific. We proved that when the numbers of antennas and users are very large, the Vanilla-GNN can only learn element-wise functions, which helps understand why the Vanilla-GNN is unable to well-learn many precoding policies. We also explained why a Vanilla-GNN for learning precoding policy cannot be generalized to the number of users by taking the ZFBF policy as an example. Based on the insight obtained from the analysis for the Vanilla-GNN and from comparing it with a perfectly trained TGNN for learning ZFBF policy, the Model-GNN was designed, where the trainable parameters play the role of adjusting the power and direction of the precoding vectors.
Simulation results validated our analyses, and showed that the Model-GNN can efficiently learn SE- and EE-maximal precoding policies, and can be generalized to the unseen scenarios with different numbers of users. Since the Model-GNN is with low training complexity, it can be re-trained with low cost for learning to optimize precoding towards different objectives and constraints simply by changing the cost function for unsupervised learning. Since the Model-GNN is generalizable to problem scales,  it can be directly used for inference in the system with a time-varying number of users without the need for re-training. The Model-GNN can also be applied for learning many other matrix inverse-related policies, such as precoding in mmWave communication and IRS systems, channel estimation and combiner in MIMO systems.

\vspace{-2mm}
\begin{appendices} \numberwithin{equation}{section}
	\renewcommand{\thesectiondis}[2]{\Alph{section}}
	\section{~~~Proof of Proposition \ref{prop: GNN element wise policy}} \label{proof: prop: GNN element wise policy} \vspace{-1mm}
		 We prove the proposition by mathematical induction. We first prove that by using the mean-pooling function, the relation between the output of the first layer and the input is an element-wise function. According to \eqref{eq: edge gnn upd1}, the output of the first layer is,\vspace{-1mm}
		\begin{equation}\label{eq: App_A-1}
		{\bf d}_{nk}^{(1)} = \textstyle\sigma\big({\bf s}h_{nk} + {\bf p}\frac{1}{N}\sum_{i=1,i\neq n}^N h_{ik} + {\bf q}\frac{1}{K}\sum_{j=1,j\neq k}^K h_{nj} \big),\vspace{-1mm}
		\end{equation}
		where ${\bf s, p}$ and ${\bf q}$ are vectors instead of matrices in \eqref{eq: edge gnn upd1} because $h_{nk}$ is a scalar. When $h_{nk}$ is i.i.d. among all the antennas and all users and $N,K\to \infty$, $\frac{1}{N}\sum_{i=1,i\neq n}^N h_{ik}\to \bar{h}, \frac{1}{K}\sum_{j=1,j\neq k}^K h_{nj} \to \bar{h}$, where $\bar{h}$ is the mean value of $h_{nk}$. Then, \eqref{eq: App_A-1} becomes
		${\bf d}_{nk}^{(1)} = \sigma({\bf s}h_{nk} + {\bf p}\bar{h} + {\bf q}\bar{h} ) \triangleq g^{(1)}(h_{nk})$,
		i.e., ${\bf d}_{nk}^{(1)}$ only depends on $h_{nk}$.
		
		We then prove that if ${\bf d}_{nk}^{(\ell-1)}$ is an element-wise function of $h_{nk}$ denoted as $g_0(\cdot)$, i.e., ${\bf d}_{nk}^{(\ell-1)}=g_0(h_{nk})$, then ${\bf d}_{nk}^{(\ell)}$ is an element-wise function of $h_{nk}$.
		
		Since ${\bf d}_{nk}^{(\ell-1)}$ is only a function of $h_{nk}$ and $\{h_{nk}, \forall n,k\}$ is independently distributed, $\{{\bf d}_{nk}^{(\ell-1)},\forall n,k\}$ is also independent with each other. The probability density function (PDF) of ${\bf d}_{nk}^{(\ell-1)}$ can be expressed as $f_{h}\big(g_0^{-1}({\bf d}_{nk}^{(\ell-1)})\big)\big|g_0^{-1'}({\bf d}_{nk}^{(\ell-1)})\big|$, where $f_h(\cdot)$ denotes the PDF of channel coefficients, and $g_0^{-1}(\cdot)$ denotes the inverse function of $g_0(\cdot)$ \cite{degroot2012probability}. This indicates that the PDF of ${\bf d}_{nk}^{(\ell-1)}$ is identical for $\forall n,k$, i.e.,  ${\bf d}_{nk}^{(\ell-1)}$ is i.i.d. for $\forall n,k$. Same with the proof that ${\bf d}_{nk}^{(1)}$ is an element-wise function of $h_{nk}$ above, we can prove that ${\bf d}_{nk}^{(\ell)}$ is an element-wise function of ${\bf d}_{nk}^{(\ell-1)}$. By using it from $\ell=1$ to $L$, we obtain $\hat{v}_{nk}={\bf d}_{nk}^{(L)}$ is an element-wise function of $h_{nk}$.
		
		The proof for the sum-pooling function is similar to the proof for the mean-pooling function, hence is omitted for conciseness. We next prove the proposition when using the max-pooling function. Again, according to \eqref{eq: edge gnn upd1}, the output of the first layer is,\vspace{-1mm}
		\begin{equation}\label{eq: App_A-3}
		{\bf d}_{nk}^{(1)} = \sigma\big({\bf s}h_{nk} + {\bf p}\max\!_{i=1,i\neq n}^N h_{ik} + {\bf q}\max\!_{j=1,j\neq k}^K h_{nj} \big).\vspace{-1mm}
		\end{equation}
		When $h_{nk}$ is i.i.d. for $\forall k,n$, and $K, N\to \infty$, $\max\!_{i=1,i\neq n}^N h_{ik}\to f_h^{-1}(\frac{N}{N+1})= f_h^{-1}(1)$, $\max\!_{j=1,j\neq k}^K h_{nj}\to f_h^{-1}(\frac{K}{K+1})= f_h^{-1}(1)$ \cite{book_order_statistics}. Then, \eqref{eq: App_A-3} becomes
		${\bf d}_{nk}^{(1)} = \sigma({\bf s}h_{nk} + {\bf q}f_h^{-1}(1) + {\bf p}f_h^{-1}(1))$,
		i.e., ${\bf d}_{nk}^{(1)}$ only depends on $h_{nk}$. Again, by using the mathematical induction method, we can prove that $\hat{v}_{nk}={\bf d}_{nk}^{(L)}$ is an element-wise function of $h_{nk}$.
		
	\vspace{-2mm}
	\section{~} \label{proof: prop: contractive mapping} \vspace{-2mm} 
		To simplify the notations, we approximate a function with scalar input $h$ and output $f(h)$ with the first-order Taylor's expansion, i.e.,
		\begin{equation}\label{eq: Taylor scalar}
		f(h) \approx f(h_0) + f'(h_0)(h-h_0).
		\end{equation}
		Denote the output of the $(\ell-1)$th iteration as $d^{(\ell-1)}$.
The input of $f(\cdot)$ that corresponds to the output $d^{(\ell-1)}$ is obtained via the inverse function of $f(\cdot)$ as $f^{-1}(d^{(\ell-1)})$.

To obtain the output of the $\ell$th iteration, we expand $f(h)$ at $f^{-1}(d^{(\ell-1)})$ via \eqref{eq: Taylor scalar}, i.e.,
			\begin{eqnarray}\label{eq: App_B-1}
			d^{(\ell)} = u(d^{(\ell-1)})&\triangleq& f(f^{-1}(d^{(\ell-1)})) + f'\big(f^{-1}(d^{(\ell-1)})\big)\big(f^{-1}(d^{(\ell-1)})-h\big) \notag\\
			&\overset{(a)}{=}& d^{(\ell-1)} + f'\big(f^{-1}(d^{(\ell-1)})\big)\big(f^{-1}(d^{(\ell-1)})-h\big),
			\end{eqnarray}
		where $(a)$ comes from $f(f^{-1}(d^{(\ell-1)}))=d^{(\ell-1)}$.
		Further considering the relationship between the derivative of $f(\cdot)$ and its inverse function $f^{-1}(\cdot)$, i.e., $f^{-1'}(\cdot)=\frac{1}{f'(\cdot)}$ \cite{book_Calculus_unlimited}, we have $f'\big(f^{-1'}(d^{(\ell-1)})\big)=1/f^{'}(d^{(\ell-1)})$. Then, \eqref{eq: App_B-1} can be re-written as,
		\begin{equation}\label{eq: App_B-2}
		d^{(\ell)} = d^{(\ell-1)} + \frac{1}{f^{-1'}(d^{(\ell-1)})}\big(f^{-1}(d^{(\ell-1)})-h\big).
		\end{equation}
		Denote $G(d)\triangleq f^{-1}(d)-h$. It is easy to obtain that $G'(d)\triangleq f^{-1'}(d)$. Then, we can re-write \eqref{eq: App_B-2} as
		$d^{(\ell)} = d^{(\ell-1)} + \frac{G(d^{(\ell-1)})}{G'(d^{(\ell-1)})}$.		
		This is exactly the iteration equation of solving $G(d)=f^{-1}(d)-h=0$ with Newton's method \cite{book_functional_analysis}. It is proved in \cite{book_functional_analysis} that the iteration equation is a contractive mapping in the neighborhood of $f(h)$ that can converge to $f^{-1}(d)-h=0$, i.e., $d=f(h)$.
		
		The proof  still holds for multivariable functions, i.e., the functions with multiple input variables and output variables. Specifically, by replacing $f(h)$ with ${\bf F}({\bf H})={\bf H}^+$, and replacing $d^{(\ell)}$ with ${\bf D}^{(\ell)}$, \eqref{eq: App_B-1} becomes \eqref{eq: pseudo inv upd}, which is a contractive mapping. According to the definition of contractive mapping \cite{book_functional_analysis}, there exists a measure of distance $\gamma(\cdot, \cdot)$ such that for an arbitrary point ${\bf D}^{(\ell)} = u({\bf D}^{(\ell-1)})$, $\gamma({\bf D}^{(\ell)}, {\bf H}^+)\leq \gamma({\bf D}^{(\ell-1)}, {\bf H}^+)$. This indicates that ${\bf D}^{(\ell)}$ is closer to ${\bf H}^+$ with more iterations.

	\vspace{-6mm}
	\section{~}\label{proof: prop: gnn scalability} \vspace{-2mm}
	In the proof, we take the sum-pooling function as an example, and the proofs for mean-pooling and max-pooling are similar. If a well-trained Vanilla-GNN can approximate the ZFBF policy accurately in the small-scale problem with $K$ users, then for arbitrary two channel matrices ${\bf H}_1=[h_{1,nk}]^{N\times K}$ and ${\bf H}_2=[h_{2,nk}]^{N\times K}$, ${\cal G}({\bf H}_1, \bm{\theta}_{\sf V}^{\star})={\bf V}_{\sf ZF}({\bf H}_1)$ and ${\cal G}({\bf H}_2, \bm{\theta}_{\sf V}^{\star})={\bf V}_{\sf ZF}({\bf H}_2)$. According to \eqref{eq: edge gnn upd}, the outputs of the well-trained Vanilla-GNN in the first layer with input  ${\bf H}_1$ and ${\bf H}_2$ are respectively,\vspace{-1mm}
	\begin{eqnarray}\label{eq: edge gnn upd opt}
	{\bf d}_{1, nk}^{(1)} &=& \sigma\Big({\bf S}^{\star}h_{1, nk} + {\bf P}^{\star}\cdot \textstyle\sum_{i=1,i\neq n}^N h_{1, ik} + {\bf Q}^{\star}\cdot\textstyle\sum_{j=1,j\neq k}^K h_{1, nj}\Big), \\
	{\bf d}_{2, nk}^{(1)} &=& \sigma\Big({\bf S}^{\star}h_{2, nk} + {\bf P}^{\star}\cdot \textstyle\sum_{i=1,i\neq n}^N h_{2, ik} + {\bf Q}^{\star}\cdot\textstyle\sum_{j=1,j\neq k}^K h_{2, nj}\Big).\vspace{-1mm}
	\end{eqnarray}
	
	In the large-scale problem with $2K$ users, the output of the  well-trained  Vanilla-GNN in the first layer with input $\bf H$ is,\vspace{-1mm}
	{\small\begin{eqnarray}\label{eq: output}
		{\bf d}_{nk}^{(1)} &=& \sigma\Big({\bf S}^{\star}h_{nk} + {\bf P}^{\star}\cdot \textstyle\sum_{i=1,i\neq n}^N h_{ik} + {\bf Q}^{\star}\cdot\textstyle\sum_{j=1,j\neq k}^{2K} h_{nj}\Big)\notag\\
		&\overset{(a)}{=}\!\!\!&\!\!\!\left\{
		\begin{aligned}
		&\sigma\Big({\bf S}^{\star}h_{1, nk} + {\bf P}^{\star}\cdot \textstyle\sum_{i=1,i\neq n}^N h_{1, ik} + {\bf Q}^{\star}\cdot\textstyle\sum_{j=1,j\neq k}^{K} h_{1, nj} + \underbrace{{\bf Q}^{\star}\cdot\textstyle\sum_{j=1}^{K} h_{2, nj}}_{(b)}\Big), k=1,\cdots,K \\
		&\sigma\Big({\bf S}^{\star}h_{2, nk'} + {\bf P}^{\star}\cdot \textstyle\sum_{i=1,i\neq n}^N h_{2, ik'} + {\bf Q}^{\star}\cdot\textstyle\sum_{j=1,j\neq k'}^{K} h_{2, nj} + \underbrace{{\bf Q}^{\star}\cdot\textstyle\sum_{j=1}^{K} h_{1, nj}}_{(c)}\Big), k=K+1,\!\cdots\!,2K,
		\end{aligned}\right.\vspace{-1mm}
		\end{eqnarray}}
	\hspace{-2mm}where $(a)$ comes from ${\bf H}=[{\bf H}_{1}, {\bf H}_{2}]$, and $k'=k-K$.

By comparing \eqref{eq: output} with \eqref{eq: edge gnn upd opt}, we can see that ${\bf d}_{nk}^{(1)}\neq {\bf d}_{1, nk}^{(1)}$ when $k=1,\cdots,K$ and ${\bf d}_{nk}^{(1)}\neq {\bf d}_{2, nk'}^{(1)}$ when $k=K+1,\cdots,2K$, because of the extra two terms $(b)$ and $(c)$  in \eqref{eq: output}.

Denote ${\bf D}_1^{(\ell)}=[{\bf d}_{1, nk}^{(\ell)}]^{N\times K}$ and ${\bf D}_2^{(\ell)}=[{\bf d}_{2, nk}^{(\ell)}]^{N\times K}$. Then, ${\bf D}^{(1)}\neq[{\bf D}_1^{(1)}, {\bf D}_2^{(1)}]$. Similarly, we can prove that ${\bf D}^{(\ell)}\neq[{\bf D}_1^{(\ell)}, {\bf D}_2^{(\ell)}]$ from $\ell=2$ to $L$. Then, the output of the Vanilla-GNN is $\hat{\bf V}\neq [{\bf D}_1^{(L)}, {\bf D}_2^{(L)}]=[{\bf V}_{\sf ZF}({\bf H}_1), {\bf V}_{\sf ZF}({\bf H}_2)]={\bf V}_{\sf ZF}({\bf H})$. This indicates that the Vanilla-GNN well-trained for the small-scale system cannot learn the ZFBF policy for the large-scale system.

\end{appendices}

\bibliography{IEEEabrv,GJ}

\end{document}